\documentclass[aps,prb,floatfix,twocolumn]{revtex4-1}

\usepackage{amssymb, amsmath, mathtools}
\usepackage{dsfont}
\DeclareMathOperator*{\argmin}{arg\,min}

\usepackage[english]{babel}
\usepackage[utf8]{inputenc}
\usepackage{blindtext}
\usepackage{graphicx}
\usepackage[colorlinks=true]{hyperref}
\usepackage{physics}
\usepackage{xcolor}
\usepackage[normalem]{ulem}
\usepackage{float}
\usepackage{pgfplots}
\usepackage{xspace}
\usepackage{footmisc}
\usepackage{placeins}
\usepackage{titlesec}

\newcommand{\pidx}[1]{{\mbox{\tiny $(#1)$}}}

\newcommand{\iden}{\mathbb{I}}
\newcommand*{\placeholderdot}{\makebox[2ex]{$\textcolor{gray}{\bullet}$}}%

\usepackage{pifont}% http://ctan.org/pkg/pifont

\newcommand{\fder}[2]{\frac{\textrm{d} #1}{\textrm{d} #2}}

\newcommand{\pSR}{p-ITE\xspace}
\newcommand{\TARNN}{ARNNO\xspace}
\newcommand{\TRBM}{RBMO\xspace}

\makeatletter
\newcommand\footnoteref[1]{\protected@xdef\@thefnmark{\ref{#1}}\@footnotemark}
\makeatother

\begin{document}

\author{Jannes Nys}
\email{Email: jannes.nys@epfl.ch}
\thanks{equal contribution.}
\affiliation{Institute of Physics, \'Ecole Polytechnique F\'ed\'erale de Lausanne (EPFL), CH-1015 Lausanne, Switzerland}
\affiliation{Center for Quantum Science and Engineering, \'Ecole Polytechnique F\'ed\'erale de Lausanne (EPFL), CH-1015 Lausanne, Switzerland}

\author{Zakari Denis}
\email{Email: zakari.denis@epfl.ch}
\thanks{equal contribution.}
\affiliation{Institute of Physics, \'Ecole Polytechnique F\'ed\'erale de Lausanne (EPFL), CH-1015 Lausanne, Switzerland}
\affiliation{Center for Quantum Science and Engineering, \'Ecole Polytechnique F\'ed\'erale de Lausanne (EPFL), CH-1015 Lausanne, Switzerland}

\author{Giuseppe Carleo}
\email{Email: giuseppe.carleo@epfl.ch}
\affiliation{Institute of Physics, \'Ecole Polytechnique F\'ed\'erale de Lausanne (EPFL), CH-1015 Lausanne, Switzerland}
\affiliation{Center for Quantum Science and Engineering, \'Ecole Polytechnique F\'ed\'erale de Lausanne (EPFL), CH-1015 Lausanne, Switzerland}

\title{Real-time quantum dynamics of thermal states with neural thermofields}

% \date{\today}

\begin{abstract}
    Solving the time-dependent quantum many-body Schr\"odinger equation is a challenging task, especially for states at a finite temperature, where the environment affects the dynamics. Most existing approximating methods are designed to represent static thermal density matrices, 1D systems, and/or zero-temperature states. In this work, we propose a method to study the real-time dynamics of thermal states in two dimensions, based on thermofield dynamics, variational Monte Carlo, and neural-network quantum states. To this aim, we introduce two novel tools: ($i$) a procedure to accurately simulate the cooling down of arbitrary quantum variational states from infinite temperature, and ($ii$) a generic thermal (autoregressive) recurrent neural-network (\TARNN) \textit{Ansatz} that allows for direct sampling from the density matrix using thermofield basis rotations. We apply our technique to the transverse-field Ising model subject to an additional longitudinal field and demonstrate that the time-dependent observables, including correlation operators, can be accurately reproduced for a $4{\times}4$ spin lattice. We provide predictions of the real-time dynamics on a $6{\times}6$ lattice that lies outside the reach of exact simulations.
\end{abstract}

\maketitle

% \tableofcontents

% \textit{Introduction.}---
\section{Introduction}
In the past decades, the understanding of thermal quantum many-body systems in contact with an environment has attracted great interest, all the more since the development of quantum (computational) technology. While major progress has been made in isolating quantum systems from the environment, its influence remains important in experimental realizations of `isolated' quantum systems. For example, near-term digital quantum simulators are intrinsically noisy (``Noisy Intermediate-Scale
Quantum" (NISQ) devices)~\cite{ebadi2021quantum, preskill2018quantum}, and therefore, there is a strong need for methods to faithfully account for the environment. 
Furthermore, analog quantum simulators such as ultracold gases in optical lattices form out-of-equilibrium experiments on many-body systems and have yielded insights into regimes of strongly correlated particles already beyond the reach of even classical simulation approaches, and will continue to challenge our understanding in the future ~\cite{gross2017quantum, brown2019bad, schafer2020tools, browaeys2020many}.
Continued progress will lead to increasing system sizes, entanglement volume, better connectivity, and/or spatial dimensions ($D>1$), which all pose significant challenges for the existing numerical toolbox, and will continue to reach far beyond the range of their applicability.

Efficiently simulating thermal systems using classical computers has been a topic of research for several decades, and various methods have been put forward (e.g. tensor-network operators)~\cite{verstraete2004, zwolak2004, kliesch2014,molnar2015,kshetrimayum2019,vanhecke2021, vanhecke2023simulatingthermaldensity}. However, it is challenging to handle the generic growth of entanglement upon time evolution in such methods~\cite{eisert2006,bravyi2007,marien2016, alhambra2021}, limiting their application to short times~\cite{osborne2006,kuwahara2021}, high temperatures~\cite{kliesch2014,molnar2015,kuwahara2021}, and/or one spatial dimension. Alternative tools to study thermal properties include quantum Monte Carlo (QMC)~\cite{barker1979quantum, shen2020finite} and the Monte Carlo wave function (MCWF) techniques~\cite{molmer1993monte,plenio1998,vicentini2019b}. More recently, thermal pure quantum (TPQ) states~\cite{sugiura2012,sugiura2013, hendry2022neural, takai2016finite} and minimally entangled typical quantum states (METTS)~\cite{white2009,stoudenmire2010,bruognolo2017}  have been introduced to approximate ensemble averages of quantum operators. 

Applications of variational Monte Carlo (VMC) to study the dynamics of thermal systems have been rather limited in the past, mainly due to the restricted expressiveness of the variational models available. In this context, the introduction of highly expressive neural quantum states (NQS)~\cite{carleo2017} opens new perspectives for faithfully capturing the full many-body structure~\cite{wu2023variational}, and the dynamics of quantum many-body problems, especially in dimension $D>1$~\cite{schmitt2020dynamics, gutierrez2022real, donatella2022, sinibaldi2023unbiasing, medvidovic2022towards}. Indeed, NQS have been shown to efficiently represent area-law and volume-law entanglement in 2D~\cite{deng2017,glasser2018,levine2019,sharir2022}. Several works have aimed at extending this construction to the density-matrix operator~\cite{torlai2018latent, vicentini2019,nagy2019,hartmann2019,yoshioka2019, nomura2021, carrasquilla2019,carrasquilla2021,luo2022, reh2021time, vicentini2022}. Thus far, dynamics of mixed states with NQS has only been demonstrated within the POVM framework, where the density matrix is not constrained to physical states~\cite{reh2021time, luo2022}. Although significant progress has been made, a general framework for accurate, Ansatz flexible, stable, consistently improvable, and faithful (i.e.\ remaining within the physical space of density matrices) real-time evolution is still lacking. Furthermore, while these previous works focus on solving the Lindblad equation, the unitary dynamics of a mixed state proves qualitatively more challenging. Indeed, the former is governed by a \textit{contractive map}, which has a fixed point. This induces a damping of numerical errors resulting in a generically stable time evolution. Moreover, the action of local Lindblad operators, as considered in the previous literature, reduces non local correlations an entanglement upon time evolution allowing for simpler Ansätze. This is in sharp contrast with unitary real-time evolution under a Hamiltonian, which is non-contractive and generically generates entanglement~\cite{eisert2006,bravyi2007,marien2016, alhambra2021}.

\begin{figure*}[htb]
    \centering
    \includegraphics[width=\textwidth]{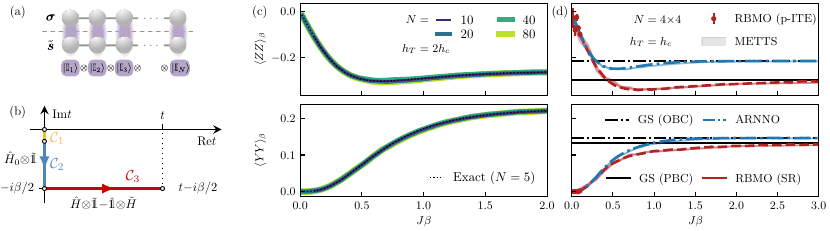}%
    \caption{(a)~Thermofield representation of the initial infinite-temperature state, corresponding to a product of locally pairwise maximally entangled Bell states. (b) Time-integration contour: this state is first evolved in imaginary time with $\hat{H}_0$ till $-i\beta/2$ (see Eq.~\eqref{eq:thermal_from_identity}), initially with an implicit solver ($\mathcal{C}_1$) and then with SR ($\mathcal{C}_2$); real-time evolution of the thermal state under $\hat{H}_t^\mathrm{th}$ (see Eq.~\eqref{eq:schrodinger}) is then performed via t-VMC ($\mathcal{C}_3$). Thermal observables of the transverse-field Ising model as a function of $\beta$ for (c)~an open chain of length $N=10, 20, 40, 80$ at $h_T=2 h_c$ with ARNNO, and (d) a $4 {\times} 4$ lattice at the critical point $h_T=h_c$. PBC were used for \TRBM, along with \pSR for the initial dynamics, whereas OBC were used for the \TARNN and plain SR thanks to our basis rotation. In (d), METTS predictions (with $3\sigma$ bands and 10k samples) and ground-state (GS) values are added for comparison. For the \TRBM, observables in the \pSR temperature regime are estimated via standard local Monte Carlo sampling and thus have higher variance than the sampling scheme used in the actual \pSR propagation. All operators are evaluated on neighboring sites.}
    \label{fig:fig1}\label{fig:N5-80_D1}\label{fig:N4_D2_crit}
\end{figure*}

In this work, we describe a variational method for performing \emph{real-time dynamics of thermal states}, bringing together NQS and the concept of thermofield dynamics~\cite{takahashi1975,suzuki1985}. Our approach relies on the preparation of an NQS in a \emph{thermal-vacuum} state through imaginary time evolution from a maximally mixed state, using a new approach to accurately time-evolve high-temperature states. We use our technique to simulate the transverse-field Ising model, subject to the sudden effect of a symmetry-breaking longitudinal field. 
Our approach allows us to study sizes beyond reach within exact simulations.

\section{Methodology}
\subsection{Thermofield dynamics}
% \textit{Thermofield dynamics.}---
Thermofield dynamics~\cite{takahashi1975,suzuki1985} is a general approach to model the behavior of thermal quantum systems, with applications in cosmology~\cite{israel1976}, condensed matter~\cite{suzuki1985} and, more recently, quantum chemistry~\cite{harsha2019,shushkov2019,harsha2019b,harsha2020a} and quantum information~\cite{chapman2019}, or to prepare thermal quantum states on quantum devices~\cite{lee2022variational, zhu2020generation}.
The underlying idea is to interpret the expectation value $\expval{\smash{\hat{O}}}_\beta$ of an observable $\hat{O}$ on a thermal state at inverse temperature $\beta = (k_B T)^{-1}$ as that on a pure state belonging to a doubled Hilbert space $\mathcal{H} \otimes \tilde{\mathcal{H}}$:
\begin{equation}
    \expval{\smash{\hat{O}}}_\beta = \frac{\Tr \bigl[\hat{O} e^{-\beta \hat{H}_0}\bigr]}{\Tr\bigl[e^{-\beta \hat{H}_0}\bigr]} = \frac{\mel{\Omega(\beta)}{\hat{O}\otimes\tilde{\mathds{1}}}{\Omega(\beta)}}{\braket{\Omega(\beta)}},\label{eq:expval_Obeta}
\end{equation}
where the ancillary space is indicated by a tilde. 
We denoted by $\tilde{O}$ an operator $\hat{O}$ acting on the thermofield space $\tilde{\mathcal{H}}$ and introduced the so-called \emph{thermal vacuum},
\begin{equation}
    \ket{\Omega(\beta)} = e^{-\frac{\beta}{2}\hat{H}_0\otimes\tilde{\mathds{1}}}\ket{\mathbb{I}},\label{eq:thermal_from_identity}
\end{equation}
corresponds to the imaginary-time evolution for a time $\beta/2$ of the \emph{identity state}. We consider a physical system consisting of $N$ spin-$1/2$ particles. In the $S_z$ basis, the identity state is expressed as
\begin{align}
    \ket{\iden} &= \bigotimes_{i=1}^N \ket{\iden_i}, \quad 
    \ket{\iden_i} = \sum_{\sigma_i, \tilde{s}_i \in \lbrace\uparrow,\downarrow\rbrace} \delta_{\sigma_i, \tilde{s}_i} \ket{\sigma_i, \tilde{s}_i}. \label{eq:identity} 
\end{align}
Here, $\vb*{\sigma}, \tilde{\vb*{s}} \in \lbrace\uparrow,\downarrow\rbrace^N$ respectively denote the physical and auxiliary (thermofield) spin-projection bases.

In Eq.~\eqref{eq:identity} the physical and thermofield degrees of freedom are pairwise maximally entangled at infinite temperature, as schematically illustrated in Fig.~\ref{fig:fig1}(a).
More generally, we refer to the thermal vacuum, or its time-evolved state, as a \emph{thermofield thermal state} $\ket{\Psi(\beta,t)}$, such that $\ket{\Psi(\beta,0)} \equiv \ket{\Omega(\beta)}$. Alternatively, the state can be denoted $\ket{\sqrt{\rho}(\beta, t)}$ since it corresponds to the purification of $\sqrt{\hat{\rho}}$. It then follows from the quadratic form of Eq.~\eqref{eq:expval_Obeta} that observable expectations are ensured to be taken on a physical (positive semi-definite) density matrix. This is in contrast, for example, with states represented in the POVM formalism~\cite{carrasquilla2019, carrasquilla2021, luo2022, reh2021time}. It is also worth noticing that, in principle, this class of operators encompasses all positive semi-definite operators, not only thermal ones.

While Eq.~\eqref{eq:thermal_from_identity} governs the imaginary-time evolution of the state of the system from the infinite-temperature state $\ket{\Psi(0,0)} = \ket{\Omega(0)} \equiv \ket{\iden}$, the real-time dynamics is translated into the following Schrödinger equation for the thermofield thermal state~\cite{gyamfi2020}:
\begin{align}
    \partial_t \ket{\Psi(\beta,t)} &= -i(\hat{H}\otimes\tilde{\mathds{1}} - \hat{\mathds{1}}\otimes\tilde{H})\ket{\Psi(\beta,t)} \nonumber \\ 
    &\equiv -i\hat{H}^\mathrm{th}\ket{\Psi(\beta,t)},\label{eq:schrodinger}
\end{align}
where the effect of tilde conjugation on operators depends upon the quantum statistics of the particles under investigation~\cite{arimitsu1985,harsha2019}. We refer to the Supplemental Material~\footnote{Supplemental Material, including references~\cite{kothe2023liouville, rackauckas2017differentialequations}.} for details on the connection to the density-matrix formalism.

\subsection{Neural thermofield states}
% \textit{Neural thermofield states.}---
In the $S_z$ basis, its neural thermofield state (NTFS) can be expressed as
\begin{equation}
    \ket{\Psi_{\vb*{\theta}}(\beta, t)} = \sum_{\vb*{\sigma},\tilde{\vb*{s}}} \psi_{\vb*{\theta}(\beta, t)}(\vb*{\sigma}, \tilde{\vb*{s}})\ket{\vb*{\sigma}, \tilde{\vb*{s}}}.\label{eq:ntfs1}
\end{equation}
Here, $\vb*{\theta}$ denote the set of variational parameters, and $\psi_{\vb*{\theta}}$ the corresponding \emph{Ansatz} wave function. In Appendix~\ref{sec:thermal_observables} we illustrate how to compute thermal observables from the NTFS in Eq.~\eqref{eq:ntfs1} using Monte Carlo sampling.

The minimal requirement for Eq.~\eqref{eq:ntfs1} to represent a viable neural thermofield state is to be able to represent the identity state \emph{exactly} for a suitable choice of initial parameters, \textit{i.e.}
\begin{equation}
    \psi_{\vb*{\theta}(\beta=0)}(\vb*{\sigma}, \tilde{\vb*{s}}) \propto \prod_i \delta_{\sigma_i,\tilde{s}_i}.
    % \psi_{\vb*{\theta}(\beta=0)}(\vb*{\sigma}, \tilde{\vb*{s}}) \propto \delta_{\vb*{\sigma}, \tilde{\vb*{s}}}.
\end{equation}
which is a product state of maximally entangled physical-thermofield spin pairs.

While such a construction is not obvious in general, we here focus on two implementations: a restricted Boltzmann machine operator (RBMO)~\cite{nomura2021, carleo2017}, and a novel autoregressive recurrent neural-network operator, which we dub ARNNO. The autoregressive property of the latter gives access to uncorrelated direct sampling from the Born distribution, dispensing from Markov-chain techniques, which has shown valuable in the study of closed quantum systems~\cite{sharir2020,hibat-allah2020,wu2022}. This remarkable property has found applications in quantum dynamics~\cite{donatella2022} and topological order~\cite{hibat2023investigating} where standard Markov-chain sampling would become challenging. A third general implementation, applicable to arbitrary neural-network architectures, is included in  Appendix~\ref{sec:architectures}, along with details about the architectures and initialization used in the numerical experiments. We next describe how to prepare the NQS in the maximally mixed state exactly and how to evolve them both in imaginary and real-time.

% \textit{Variational thermal-state preparation}--- 
\subsection{Variational thermal-state preparation}
The overall procedure for carrying out the real-time evolution of a thermal state in the thermofield picture is shown in Fig.~\ref{fig:fig1}(b). In both real and imaginary time, one can use the time-dependent variational principle (TDVP)~\cite{yuan2019, broeckhove1988equivalence}. In conjunction with Monte Carlo estimates for the expectation values, imaginary-time TDVP reduces to stochastic reconfiguration (SR)~\cite{sorella1998green}.
A major obstacle prevents one from preparing the thermal-vacuum state via SR ($\mathcal{C}_1$). Indeed, in the $S_z$ basis, the gradients of the expectation value of the time-propagator (also called ``forces''~\cite{vicentini2022netket}) vanish at $\beta = 0$ when Monte Carlo estimates are used. On the other hand, current implicit methods~\cite{sinibaldi2023unbiasing} have fidelity gradients that are dominated by probability amplitudes with large magnitudes, and, therefore, cannot depart from the infinite-temperature state. To circumvent these issues, it was proposed to add noise to the parameters of the \textit{Ansatz}~\cite{nomura2021} at the cost of introducing \emph{uncontrollable errors and bias} in the imaginary time evolution, eventually rendering the method unreliable in studying real-time evolution.
We propose two solutions to this obstacle. 
Our first approach is based on the observation that the above-mentioned bias occurs due to the presence of many zeros in the maximally-entangled identity state. The zeros in the wave function can be lifted by applying a Hadamard rotation to the auxiliary thermofield doubled spins: $\left(\ket{00}+\ket{11}\right)^{\otimes N} \to \left(\ket{00}+\ket{01}+ \ket{10} - \ket{11}\right)^{\otimes N}$. The rotated basis resolves the bias such that standard SR can be used to depart from $\beta=0$ in $\mathcal{C}_1$.
Our second approach is to introduce a novel implicit variational method dubbed projected Imaginary Time Evolution (\pSR), which we apply to start the imaginary-time evolution in $\mathcal{C}_1$ till the gradient signal can be accurately resolved. 
Therefore, we resort to an implicit formulation of TDVP~\cite{gutierrez2022real, sinibaldi2023unbiasing} and enforce a finite sample density on zero-amplitude states responsible for biasing the forces by means of self-normalized importance sampling using the following suitable prior distribution from which we can sample directly:
\begin{equation}
    q(\vb*{S}) =\mathcal{N} \sum_{\vb*{S}'} \lvert\braket{\vb*{S}'}{\iden}\rvert^2 f\bigl(\vb*{S}', \vb*{S}\bigr),
\end{equation}
where $\vb*{S} = (\vb*{\sigma}, \tilde{\vb*{s}})$, $\mathcal{N}$ is a normalization constant, and $f$ is a suitably chosen convolutional kernel. Details about the estimators are given in Appendix~\ref{sec:pite}. %Supplemental Material~\cite{Note1}.

% \textit{Neural thermofield dynamics: real-time evolution}---
\subsection{Neural thermofield dynamics: real-time evolution}
For section $\mathcal{C}_3$ in Fig.~\ref{fig:fig1}(b), we can now carry out the real-time evolution as if we were handling a closed quantum system in the enlarged Hilbert space, subject to the thermofield Hamiltonian of Eq.~\eqref{eq:schrodinger}. In conjunction with Monte Carlo, TDVP in this case reduces to the time-dependent Variational Monte Carlo (t-VMC)~\cite{carleo_localization_2012,carleo2017unitary, nys2024ab}.

\begin{figure*}[th!]
    \centering
    \includegraphics[width=.85\textwidth]{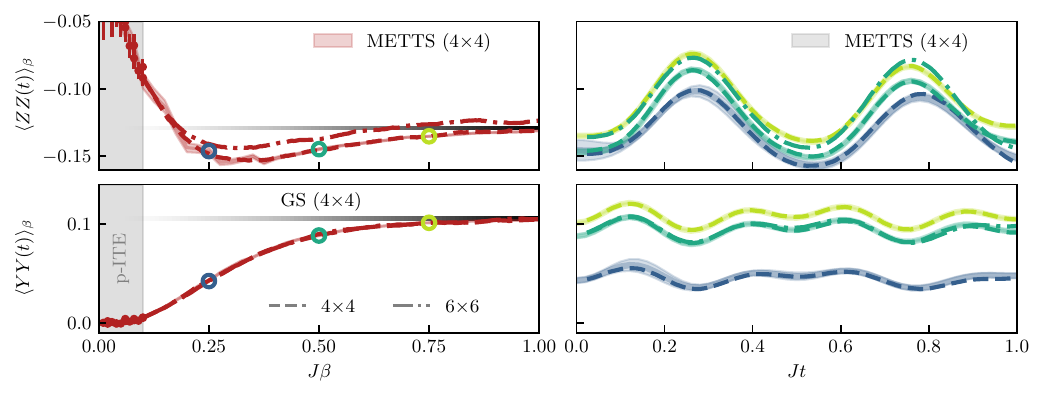}
    \caption{Thermal observables evolution as a function of $\beta$ (left) and real time $t$ (right) for a $4{\times}4$ spin lattice. METTS predictions (with $3\sigma$ bands and 10k samples) and ground-state (GS) values ($\beta \to \infty$) are added for comparison. We also show the results at $\beta=1/2$ for a $6{\times}6$ lattice, which are hard to distinguish from the $4{\times}4$ results for the $YY$ operator. For the \TRBM, observables in the \pSR temperature regime are estimated via standard local Monte Carlo sampling and thus have higher variance than the sampling scheme used in the actual \pSR propagation. Operators are evaluated on neighboring sites.}
    \label{fig:N4_D2}
\end{figure*}

\section{Results}
% \textit{Transverse- and longitudinal-field Ising model.}---
We first consider the generalized transverse-field Ising model, as described by the Hamiltonian
\begin{align}
    \hat{H}(h_T, h_L) &= J \sum_{\left<i, j\right>} \hat{\sigma}_i^{z} \hat{\sigma}_j^{z} - h_T \sum_i \hat{\sigma}_i^x + h_L \sum_i \hat{\sigma}_i^z,
\end{align}
where $J$ denotes the nearest-neighbor interaction strength, and $h_T$ and $h_L$ are the transverse and longitudinal fields, respectively. At zero temperature and zero longitudinal field, this system exhibits a second-order phase transition at $h_c = J$ in 1D and $h_c = 3.04438(2)J$ in 2D~\cite{bloete2002}.

We simulate the thermal states for chains of length up to $N=80$ at $\hat{H}_0 = \hat{H}(2h_c, 0)$ with ARNNO in Fig.~\ref{fig:fig1}(c), and for a $4{\times}4$ lattice with the critical Hamiltonian $\hat{H}_0 = \hat{H}(h_c, 0)$ with either architecture in Fig.~\ref{fig:fig1}(d), thus demonstrating the scalability and accuracy of our technique for thermal-state preparation, even at the critical point. In Fig.~\ref{fig:fig1}(d) we compare our simulations against exact METTS predictions~\footnote{Notice also that mapping the real-time dynamics of a quantum system in the METTS/TPQ formalism is challenging with variational methods~\cite{hendry2022neural} (since an ensemble of time-evolving states must be traced) for small and intermediate system sizes within reach of numerical simulations, where many samples must be propagated.}
% for a 2D system using both the \TARNN and \TRBM thermal states 
(see Supplemental Material~\cite{Note1} for further details). 

At the critical point, the system is known to exhibit singular magnetic susceptibility in the thermodynamic limit along the longitudinal direction~\cite{stinchcombe1973}. We probe the dynamics in this direction through the $ZZ$ operator. This quantifies the density of topological defects in the system and is known as a meaningful probe to interesting critical universal phenomena such as the Kibble-Zurek mechanism~\cite{puebla2019,donatella2022}. We also measure the $YY$ operator, showing the technique also proves accurate when dealing with off-diagonal operators.
In both cases, the difference with METTS is hardly visible, thereby validating the accuracy of our method in reconstructing thermal states over a wide range of temperatures, even when the underlying density matrices are non-trivial. In particular, one observes that the \textit{Ansatz} correctly interpolates between the infinite-T maximally mixed state and the closed-system ground-state asymptotics.

Next, we carry out the real-time evolution of a thermal state. To this aim, we first prepare the thermal state of the Hamiltonian $\hat{H}_0 = \hat{H}(1.5h_c, 0)$ at several temperatures of interest. The system is subsequently evolved according to $\hat{H}_t = \hat{H}(1.5h_c, J)$, simulating the sudden switching on of a parity-breaking longitudinal field. The latter is obtained by evolving the finite-temperature (doubled) state with the thermofield Hamiltonian $\hat{H}_t^\mathrm{th}$ defined in accordance with Eq.~\eqref{eq:schrodinger}. In Fig.~\ref{fig:N4_D2}, we show the results for 2D lattices of size $4{\times} 4$ and $6{\times}6$, respectively (see Appendix~\ref{sec:chain_benchmark} for simulations on a 1D chain of $10$ spins, and a comparison with exact diagonalization). The validity of the $6{\times}6$ results are demonstrated in Appendix~\ref{sec:comparison_4x4_and_6x6}. Our simulations faithfully reproduce the oscillations induced by the longitudinal external field onto the system, originally in an orthogonally polarized paramagnetic phase.
For these experiments, we use the \TRBM architecture since it is holomorphic and therefore results in stable time evolution with tVMC. For \TARNN, implicit time evolution techniques such as the one introduced in Ref.~\cite{sinibaldi2023unbiasing} prove valuable. However, in certain cases, recurrent autoregressive models can be limited in expressive power~\cite{lin2022scaling}. Nevertheless, they can prove powerful whenever Markov-chain sampling becomes challenging~\cite{donatella2022}. We also show predictions from METTS and the corresponding variance. We conclude that our method can reliably capture real-time dynamics for a wide range of temperatures. Especially as the temperature decreases, where our results become increasingly accurate since the variance on the Monte Carlo estimators in tVMC decreases, thereby requiring fewer samples to obtain a similar accuracy in the energy gradients. We describe the scalability of our method in the Supplemental Material~\cite{Note1}.

% \textit{Discussion and outlook.}---
\section{Discussion and outlook}
We introduced the framework of thermofield dynamics to capture the real-time evolution of thermal ensembles using neural network quantum states. We demonstrated its accuracy and scalability on the Ising model with a longitudinal and transverse field. First, we solve the problem of accurately preparing a neural density operator at a given temperature. We do this by introducing both novel autoregressive neural network operators (\TARNN) and a stable implicit imaginary-time evolution technique (\pSR) that allows one to cool down generic neural density operators. Unlike the POVM-based formalism, our variational thermal states are guaranteed to be positive semi-definite, and, therefore, physical. Within the thermofield formalism, we perform real-time evolution of these thermal neural density operators and can scale our simulations beyond system sizes accessible with exact methods.
Since the generality of our approach, in principle, allows one to build neural density operators from any neural quantum architecture, we foresee many extensions. This should prove useful in simulating electronic systems~\cite{nys2022variational}, with application in quantum chemistry, material science, and condensed matter; and may result in a better understanding of temperature dependence of analog quantum simulators and the influence of environmental noise on digital quantum devices.

\begin{acknowledgments}
We thank Alessandro Sinibaldi for many useful discussions. This work was supported by Microsoft Research, and by SEFRI under Grant No.\ MB22.00051 (NEQS - Neural Quantum Simulation). All simulations were carried out using NetKet~\cite{vicentini2022netket} and Jax~\cite{jax2018github}. Independent Markov-chain samplers were parallelized with MPI4Jax~\cite{hafner2021mpi4jax}. 
\end{acknowledgments}

\bibliography{biblio}

\begin{thebibliography}{89}%
\makeatletter
\providecommand \@ifxundefined [1]{%
 \@ifx{#1\undefined}
}%
\providecommand \@ifnum [1]{%
 \ifnum #1\expandafter \@firstoftwo
 \else \expandafter \@secondoftwo
 \fi
}%
\providecommand \@ifx [1]{%
 \ifx #1\expandafter \@firstoftwo
 \else \expandafter \@secondoftwo
 \fi
}%
\providecommand \natexlab [1]{#1}%
\providecommand \enquote  [1]{``#1''}%
\providecommand \bibnamefont  [1]{#1}%
\providecommand \bibfnamefont [1]{#1}%
\providecommand \citenamefont [1]{#1}%
\providecommand \href@noop [0]{\@secondoftwo}%
\providecommand \href [0]{\begingroup \@sanitize@url \@href}%
\providecommand \@href[1]{\@@startlink{#1}\@@href}%
\providecommand \@@href[1]{\endgroup#1\@@endlink}%
\providecommand \@sanitize@url [0]{\catcode `\\12\catcode `\$12\catcode
  `\&12\catcode `\#12\catcode `\^12\catcode `\_12\catcode `\%12\relax}%
\providecommand \@@startlink[1]{}%
\providecommand \@@endlink[0]{}%
\providecommand \url  [0]{\begingroup\@sanitize@url \@url }%
\providecommand \@url [1]{\endgroup\@href {#1}{\urlprefix }}%
\providecommand \urlprefix  [0]{URL }%
\providecommand \Eprint [0]{\href }%
\providecommand \doibase [0]{http://dx.doi.org/}%
\providecommand \selectlanguage [0]{\@gobble}%
\providecommand \bibinfo  [0]{\@secondoftwo}%
\providecommand \bibfield  [0]{\@secondoftwo}%
\providecommand \translation [1]{[#1]}%
\providecommand \BibitemOpen [0]{}%
\providecommand \bibitemStop [0]{}%
\providecommand \bibitemNoStop [0]{.\EOS\space}%
\providecommand \EOS [0]{\spacefactor3000\relax}%
\providecommand \BibitemShut  [1]{\csname bibitem#1\endcsname}%
\let\auto@bib@innerbib\@empty
%</preamble>
\bibitem [{\citenamefont {Ebadi}\ \emph {et~al.}(2021)\citenamefont {Ebadi},
  \citenamefont {Wang}, \citenamefont {Levine}, \citenamefont {Keesling},
  \citenamefont {Semeghini}, \citenamefont {Omran}, \citenamefont {Bluvstein},
  \citenamefont {Samajdar}, \citenamefont {Pichler}, \citenamefont {Ho} \emph
  {et~al.}}]{ebadi2021quantum}%
  \BibitemOpen
  \bibfield  {author} {\bibinfo {author} {\bibfnamefont {S.}~\bibnamefont
  {Ebadi}}, \bibinfo {author} {\bibfnamefont {T.~T.}\ \bibnamefont {Wang}},
  \bibinfo {author} {\bibfnamefont {H.}~\bibnamefont {Levine}}, \bibinfo
  {author} {\bibfnamefont {A.}~\bibnamefont {Keesling}}, \bibinfo {author}
  {\bibfnamefont {G.}~\bibnamefont {Semeghini}}, \bibinfo {author}
  {\bibfnamefont {A.}~\bibnamefont {Omran}}, \bibinfo {author} {\bibfnamefont
  {D.}~\bibnamefont {Bluvstein}}, \bibinfo {author} {\bibfnamefont
  {R.}~\bibnamefont {Samajdar}}, \bibinfo {author} {\bibfnamefont
  {H.}~\bibnamefont {Pichler}}, \bibinfo {author} {\bibfnamefont {W.~W.}\
  \bibnamefont {Ho}},  \emph {et~al.},\ }\href {\doibase
  10.1038/s41586-021-03582-4} {\bibfield  {journal} {\bibinfo  {journal}
  {Nature}\ }\textbf {\bibinfo {volume} {595}},\ \bibinfo {pages} {227}
  (\bibinfo {year} {2021})}\BibitemShut {NoStop}%
\bibitem [{\citenamefont {Preskill}(2018)}]{preskill2018quantum}%
  \BibitemOpen
  \bibfield  {author} {\bibinfo {author} {\bibfnamefont {J.}~\bibnamefont
  {Preskill}},\ }\href {\doibase 10.22331/q-2018-08-06-79} {\bibfield
  {journal} {\bibinfo  {journal} {Quantum}\ }\textbf {\bibinfo {volume} {2}},\
  \bibinfo {pages} {79} (\bibinfo {year} {2018})}\BibitemShut {NoStop}%
\bibitem [{\citenamefont {Gross}\ and\ \citenamefont
  {Bloch}(2017)}]{gross2017quantum}%
  \BibitemOpen
  \bibfield  {author} {\bibinfo {author} {\bibfnamefont {C.}~\bibnamefont
  {Gross}}\ and\ \bibinfo {author} {\bibfnamefont {I.}~\bibnamefont {Bloch}},\
  }\href {\doibase 10.1126/science.aal3837} {\bibfield  {journal} {\bibinfo
  {journal} {Science}\ }\textbf {\bibinfo {volume} {357}},\ \bibinfo {pages}
  {995} (\bibinfo {year} {2017})}\BibitemShut {NoStop}%
\bibitem [{\citenamefont {Brown}\ \emph {et~al.}(2019)\citenamefont {Brown},
  \citenamefont {Mitra}, \citenamefont {Guardado-Sanchez}, \citenamefont
  {Nourafkan}, \citenamefont {Reymbaut}, \citenamefont {Hébert}, \citenamefont
  {Bergeron}, \citenamefont {Tremblay}, \citenamefont {Kokalj}, \citenamefont
  {Huse}, \citenamefont {Schauß},\ and\ \citenamefont {Bakr}}]{brown2019bad}%
  \BibitemOpen
  \bibfield  {author} {\bibinfo {author} {\bibfnamefont {P.~T.}\ \bibnamefont
  {Brown}}, \bibinfo {author} {\bibfnamefont {D.}~\bibnamefont {Mitra}},
  \bibinfo {author} {\bibfnamefont {E.}~\bibnamefont {Guardado-Sanchez}},
  \bibinfo {author} {\bibfnamefont {R.}~\bibnamefont {Nourafkan}}, \bibinfo
  {author} {\bibfnamefont {A.}~\bibnamefont {Reymbaut}}, \bibinfo {author}
  {\bibfnamefont {C.-D.}\ \bibnamefont {Hébert}}, \bibinfo {author}
  {\bibfnamefont {S.}~\bibnamefont {Bergeron}}, \bibinfo {author}
  {\bibfnamefont {A.-M.~S.}\ \bibnamefont {Tremblay}}, \bibinfo {author}
  {\bibfnamefont {J.}~\bibnamefont {Kokalj}}, \bibinfo {author} {\bibfnamefont
  {D.~A.}\ \bibnamefont {Huse}}, \bibinfo {author} {\bibfnamefont
  {P.}~\bibnamefont {Schauß}}, \ and\ \bibinfo {author} {\bibfnamefont
  {W.~S.}\ \bibnamefont {Bakr}},\ }\href {\doibase 10.1126/science.aat4134}
  {\bibfield  {journal} {\bibinfo  {journal} {Science}\ }\textbf {\bibinfo
  {volume} {363}},\ \bibinfo {pages} {379} (\bibinfo {year}
  {2019})}\BibitemShut {NoStop}%
\bibitem [{\citenamefont {Sch{\"a}fer}\ \emph {et~al.}(2020)\citenamefont
  {Sch{\"a}fer}, \citenamefont {Fukuhara}, \citenamefont {Sugawa},
  \citenamefont {Takasu},\ and\ \citenamefont {Takahashi}}]{schafer2020tools}%
  \BibitemOpen
  \bibfield  {author} {\bibinfo {author} {\bibfnamefont {F.}~\bibnamefont
  {Sch{\"a}fer}}, \bibinfo {author} {\bibfnamefont {T.}~\bibnamefont
  {Fukuhara}}, \bibinfo {author} {\bibfnamefont {S.}~\bibnamefont {Sugawa}},
  \bibinfo {author} {\bibfnamefont {Y.}~\bibnamefont {Takasu}}, \ and\ \bibinfo
  {author} {\bibfnamefont {Y.}~\bibnamefont {Takahashi}},\ }\href {\doibase
  10.1038/s42254-020-0195-3} {\bibfield  {journal} {\bibinfo  {journal} {Nature
  Reviews Physics}\ }\textbf {\bibinfo {volume} {2}},\ \bibinfo {pages} {411}
  (\bibinfo {year} {2020})}\BibitemShut {NoStop}%
\bibitem [{\citenamefont {Browaeys}\ and\ \citenamefont
  {Lahaye}(2020)}]{browaeys2020many}%
  \BibitemOpen
  \bibfield  {author} {\bibinfo {author} {\bibfnamefont {A.}~\bibnamefont
  {Browaeys}}\ and\ \bibinfo {author} {\bibfnamefont {T.}~\bibnamefont
  {Lahaye}},\ }\href {\doibase 10.1038/s41567-019-0733-z} {\bibfield  {journal}
  {\bibinfo  {journal} {Nature Physics}\ }\textbf {\bibinfo {volume} {16}},\
  \bibinfo {pages} {132} (\bibinfo {year} {2020})}\BibitemShut {NoStop}%
\bibitem [{\citenamefont {Verstraete}\ \emph {et~al.}(2004)\citenamefont
  {Verstraete}, \citenamefont {{Garc{\'i}a-Ripoll}},\ and\ \citenamefont
  {Cirac}}]{verstraete2004}%
  \BibitemOpen
  \bibfield  {author} {\bibinfo {author} {\bibfnamefont {F.}~\bibnamefont
  {Verstraete}}, \bibinfo {author} {\bibfnamefont {J.~J.}\ \bibnamefont
  {{Garc{\'i}a-Ripoll}}}, \ and\ \bibinfo {author} {\bibfnamefont {J.~I.}\
  \bibnamefont {Cirac}},\ }\href {\doibase 10.1103/PhysRevLett.93.207204}
  {\bibfield  {journal} {\bibinfo  {journal} {Physical Review Letters}\
  }\textbf {\bibinfo {volume} {93}},\ \bibinfo {pages} {207204} (\bibinfo
  {year} {2004})}\BibitemShut {NoStop}%
\bibitem [{\citenamefont {Zwolak}\ and\ \citenamefont
  {Vidal}(2004)}]{zwolak2004}%
  \BibitemOpen
  \bibfield  {author} {\bibinfo {author} {\bibfnamefont {M.}~\bibnamefont
  {Zwolak}}\ and\ \bibinfo {author} {\bibfnamefont {G.}~\bibnamefont {Vidal}},\
  }\href {\doibase 10.1103/PhysRevLett.93.207205} {\bibfield  {journal}
  {\bibinfo  {journal} {Physical Review Letters}\ }\textbf {\bibinfo {volume}
  {93}},\ \bibinfo {pages} {207205} (\bibinfo {year} {2004})}\BibitemShut
  {NoStop}%
\bibitem [{\citenamefont {Kliesch}\ \emph {et~al.}(2014)\citenamefont
  {Kliesch}, \citenamefont {Gogolin}, \citenamefont {Kastoryano}, \citenamefont
  {Riera},\ and\ \citenamefont {Eisert}}]{kliesch2014}%
  \BibitemOpen
  \bibfield  {author} {\bibinfo {author} {\bibfnamefont {M.}~\bibnamefont
  {Kliesch}}, \bibinfo {author} {\bibfnamefont {C.}~\bibnamefont {Gogolin}},
  \bibinfo {author} {\bibfnamefont {M.~J.}\ \bibnamefont {Kastoryano}},
  \bibinfo {author} {\bibfnamefont {A.}~\bibnamefont {Riera}}, \ and\ \bibinfo
  {author} {\bibfnamefont {J.}~\bibnamefont {Eisert}},\ }\href {\doibase
  10.1103/PhysRevX.4.031019} {\bibfield  {journal} {\bibinfo  {journal}
  {Physical Review X}\ }\textbf {\bibinfo {volume} {4}},\ \bibinfo {pages}
  {031019} (\bibinfo {year} {2014})}\BibitemShut {NoStop}%
\bibitem [{\citenamefont {Molnar}\ \emph {et~al.}(2015)\citenamefont {Molnar},
  \citenamefont {Schuch}, \citenamefont {Verstraete},\ and\ \citenamefont
  {Cirac}}]{molnar2015}%
  \BibitemOpen
  \bibfield  {author} {\bibinfo {author} {\bibfnamefont {A.}~\bibnamefont
  {Molnar}}, \bibinfo {author} {\bibfnamefont {N.}~\bibnamefont {Schuch}},
  \bibinfo {author} {\bibfnamefont {F.}~\bibnamefont {Verstraete}}, \ and\
  \bibinfo {author} {\bibfnamefont {J.~I.}\ \bibnamefont {Cirac}},\ }\href
  {\doibase 10.1103/PhysRevB.91.045138} {\bibfield  {journal} {\bibinfo
  {journal} {Physical Review B}\ }\textbf {\bibinfo {volume} {91}},\ \bibinfo
  {pages} {045138} (\bibinfo {year} {2015})}\BibitemShut {NoStop}%
\bibitem [{\citenamefont {Kshetrimayum}\ \emph {et~al.}(2019)\citenamefont
  {Kshetrimayum}, \citenamefont {Rizzi}, \citenamefont {Eisert},\ and\
  \citenamefont {Or{\'u}s}}]{kshetrimayum2019}%
  \BibitemOpen
  \bibfield  {author} {\bibinfo {author} {\bibfnamefont {A.}~\bibnamefont
  {Kshetrimayum}}, \bibinfo {author} {\bibfnamefont {M.}~\bibnamefont {Rizzi}},
  \bibinfo {author} {\bibfnamefont {J.}~\bibnamefont {Eisert}}, \ and\ \bibinfo
  {author} {\bibfnamefont {R.}~\bibnamefont {Or{\'u}s}},\ }\href {\doibase
  10.1103/PhysRevLett.122.070502} {\bibfield  {journal} {\bibinfo  {journal}
  {Physical Review Letters}\ }\textbf {\bibinfo {volume} {122}},\ \bibinfo
  {pages} {070502} (\bibinfo {year} {2019})}\BibitemShut {NoStop}%
\bibitem [{\citenamefont {Vanhecke}\ \emph {et~al.}(2021)\citenamefont
  {Vanhecke}, \citenamefont {Devoogdt}, \citenamefont {Verstraete},\ and\
  \citenamefont {Vanderstraeten}}]{vanhecke2021}%
  \BibitemOpen
  \bibfield  {author} {\bibinfo {author} {\bibfnamefont {B.}~\bibnamefont
  {Vanhecke}}, \bibinfo {author} {\bibfnamefont {D.}~\bibnamefont {Devoogdt}},
  \bibinfo {author} {\bibfnamefont {F.}~\bibnamefont {Verstraete}}, \ and\
  \bibinfo {author} {\bibfnamefont {L.}~\bibnamefont {Vanderstraeten}},\ }\href
  {\doibase 10.48550/arXiv.2112.01507} {\enquote {\bibinfo {title} {Simulating
  thermal density operators with cluster expansions and tensor networks},}\ }
  (\bibinfo {year} {2021}),\ \Eprint {http://arxiv.org/abs/2112.01507}
  {arXiv:2112.01507} \BibitemShut {NoStop}%
\bibitem [{\citenamefont {Vanhecke}\ \emph {et~al.}(2023)\citenamefont
  {Vanhecke}, \citenamefont {Devoogdt}, \citenamefont {Verstraete},\ and\
  \citenamefont {Vanderstraeten}}]{vanhecke2023simulatingthermaldensity}%
  \BibitemOpen
  \bibfield  {author} {\bibinfo {author} {\bibfnamefont {B.}~\bibnamefont
  {Vanhecke}}, \bibinfo {author} {\bibfnamefont {D.}~\bibnamefont {Devoogdt}},
  \bibinfo {author} {\bibfnamefont {F.}~\bibnamefont {Verstraete}}, \ and\
  \bibinfo {author} {\bibfnamefont {L.}~\bibnamefont {Vanderstraeten}},\ }\href
  {\doibase 10.21468/SciPostPhys.14.4.085} {\bibfield  {journal} {\bibinfo
  {journal} {SciPost Phys.}\ }\textbf {\bibinfo {volume} {14}},\ \bibinfo
  {pages} {085} (\bibinfo {year} {2023})}\BibitemShut {NoStop}%
\bibitem [{\citenamefont {Eisert}\ and\ \citenamefont
  {Osborne}(2006)}]{eisert2006}%
  \BibitemOpen
  \bibfield  {author} {\bibinfo {author} {\bibfnamefont {J.}~\bibnamefont
  {Eisert}}\ and\ \bibinfo {author} {\bibfnamefont {T.~J.}\ \bibnamefont
  {Osborne}},\ }\href {\doibase 10.1103/PhysRevLett.97.150404} {\bibfield
  {journal} {\bibinfo  {journal} {Physical Review Letters}\ }\textbf {\bibinfo
  {volume} {97}},\ \bibinfo {pages} {150404} (\bibinfo {year}
  {2006})}\BibitemShut {NoStop}%
\bibitem [{\citenamefont {Bravyi}(2007)}]{bravyi2007}%
  \BibitemOpen
  \bibfield  {author} {\bibinfo {author} {\bibfnamefont {S.}~\bibnamefont
  {Bravyi}},\ }\href {\doibase 10.1103/PhysRevA.76.052319} {\bibfield
  {journal} {\bibinfo  {journal} {Physical Review A}\ }\textbf {\bibinfo
  {volume} {76}},\ \bibinfo {pages} {052319} (\bibinfo {year}
  {2007})}\BibitemShut {NoStop}%
\bibitem [{\citenamefont {Mari{\"e}n}\ \emph {et~al.}(2016)\citenamefont
  {Mari{\"e}n}, \citenamefont {Audenaert}, \citenamefont {Van~Acoleyen},\ and\
  \citenamefont {Verstraete}}]{marien2016}%
  \BibitemOpen
  \bibfield  {author} {\bibinfo {author} {\bibfnamefont {M.}~\bibnamefont
  {Mari{\"e}n}}, \bibinfo {author} {\bibfnamefont {K.~M.~R.}\ \bibnamefont
  {Audenaert}}, \bibinfo {author} {\bibfnamefont {K.}~\bibnamefont
  {Van~Acoleyen}}, \ and\ \bibinfo {author} {\bibfnamefont {F.}~\bibnamefont
  {Verstraete}},\ }\href {\doibase 10.1007/s00220-016-2709-5} {\bibfield
  {journal} {\bibinfo  {journal} {Communications in Mathematical Physics}\
  }\textbf {\bibinfo {volume} {346}},\ \bibinfo {pages} {35} (\bibinfo {year}
  {2016})}\BibitemShut {NoStop}%
\bibitem [{\citenamefont {Alhambra}\ and\ \citenamefont
  {Cirac}(2021)}]{alhambra2021}%
  \BibitemOpen
  \bibfield  {author} {\bibinfo {author} {\bibfnamefont {{\'A}.~M.}\
  \bibnamefont {Alhambra}}\ and\ \bibinfo {author} {\bibfnamefont {J.~I.}\
  \bibnamefont {Cirac}},\ }\href {\doibase 10.1103/PRXQuantum.2.040331}
  {\bibfield  {journal} {\bibinfo  {journal} {PRX Quantum}\ }\textbf {\bibinfo
  {volume} {2}},\ \bibinfo {pages} {040331} (\bibinfo {year}
  {2021})}\BibitemShut {NoStop}%
\bibitem [{\citenamefont {Osborne}(2006)}]{osborne2006}%
  \BibitemOpen
  \bibfield  {author} {\bibinfo {author} {\bibfnamefont {T.~J.}\ \bibnamefont
  {Osborne}},\ }\href {\doibase 10.1103/PhysRevLett.97.157202} {\bibfield
  {journal} {\bibinfo  {journal} {Physical Review Letters}\ }\textbf {\bibinfo
  {volume} {97}},\ \bibinfo {pages} {157202} (\bibinfo {year}
  {2006})}\BibitemShut {NoStop}%
\bibitem [{\citenamefont {Kuwahara}\ \emph {et~al.}(2021)\citenamefont
  {Kuwahara}, \citenamefont {Alhambra},\ and\ \citenamefont
  {Anshu}}]{kuwahara2021}%
  \BibitemOpen
  \bibfield  {author} {\bibinfo {author} {\bibfnamefont {T.}~\bibnamefont
  {Kuwahara}}, \bibinfo {author} {\bibfnamefont {{\'A}.~M.}\ \bibnamefont
  {Alhambra}}, \ and\ \bibinfo {author} {\bibfnamefont {A.}~\bibnamefont
  {Anshu}},\ }\href {\doibase 10.1103/PhysRevX.11.011047} {\bibfield  {journal}
  {\bibinfo  {journal} {Physical Review X}\ }\textbf {\bibinfo {volume} {11}},\
  \bibinfo {pages} {011047} (\bibinfo {year} {2021})}\BibitemShut {NoStop}%
\bibitem [{\citenamefont {Barker}(2008)}]{barker1979quantum}%
  \BibitemOpen
  \bibfield  {author} {\bibinfo {author} {\bibfnamefont {J.~A.}\ \bibnamefont
  {Barker}},\ }\href {\doibase 10.1063/1.437829} {\bibfield  {journal}
  {\bibinfo  {journal} {The Journal of Chemical Physics}\ }\textbf {\bibinfo
  {volume} {70}},\ \bibinfo {pages} {2914} (\bibinfo {year}
  {2008})}\BibitemShut {NoStop}%
\bibitem [{\citenamefont {Shen}\ \emph {et~al.}(2020)\citenamefont {Shen},
  \citenamefont {Liu}, \citenamefont {Yu},\ and\ \citenamefont
  {Rubenstein}}]{shen2020finite}%
  \BibitemOpen
  \bibfield  {author} {\bibinfo {author} {\bibfnamefont {T.}~\bibnamefont
  {Shen}}, \bibinfo {author} {\bibfnamefont {Y.}~\bibnamefont {Liu}}, \bibinfo
  {author} {\bibfnamefont {Y.}~\bibnamefont {Yu}}, \ and\ \bibinfo {author}
  {\bibfnamefont {B.~M.}\ \bibnamefont {Rubenstein}},\ }\href {\doibase
  10.1063/5.0026606} {\bibfield  {journal} {\bibinfo  {journal} {The Journal of
  Chemical Physics}\ }\textbf {\bibinfo {volume} {153}},\ \bibinfo {pages}
  {204108} (\bibinfo {year} {2020})}\BibitemShut {NoStop}%
\bibitem [{\citenamefont {M{\o}lmer}\ \emph {et~al.}(1993)\citenamefont
  {M{\o}lmer}, \citenamefont {Castin},\ and\ \citenamefont
  {Dalibard}}]{molmer1993monte}%
  \BibitemOpen
  \bibfield  {author} {\bibinfo {author} {\bibfnamefont {K.}~\bibnamefont
  {M{\o}lmer}}, \bibinfo {author} {\bibfnamefont {Y.}~\bibnamefont {Castin}}, \
  and\ \bibinfo {author} {\bibfnamefont {J.}~\bibnamefont {Dalibard}},\ }\href
  {\doibase 10.1364/JOSAB.10.000524} {\bibfield  {journal} {\bibinfo  {journal}
  {J. Opt. Soc. Am. B}\ }\textbf {\bibinfo {volume} {10}},\ \bibinfo {pages}
  {524} (\bibinfo {year} {1993})}\BibitemShut {NoStop}%
\bibitem [{\citenamefont {Plenio}\ and\ \citenamefont
  {Knight}(1998)}]{plenio1998}%
  \BibitemOpen
  \bibfield  {author} {\bibinfo {author} {\bibfnamefont {M.~B.}\ \bibnamefont
  {Plenio}}\ and\ \bibinfo {author} {\bibfnamefont {P.~L.}\ \bibnamefont
  {Knight}},\ }\href {\doibase 10.1103/RevModPhys.70.101} {\bibfield  {journal}
  {\bibinfo  {journal} {Rev. Mod. Phys.}\ }\textbf {\bibinfo {volume} {70}},\
  \bibinfo {pages} {101} (\bibinfo {year} {1998})}\BibitemShut {NoStop}%
\bibitem [{\citenamefont {Vicentini}\ \emph
  {et~al.}(2019{\natexlab{a}})\citenamefont {Vicentini}, \citenamefont
  {Minganti}, \citenamefont {Biella}, \citenamefont {Orso},\ and\ \citenamefont
  {Ciuti}}]{vicentini2019b}%
  \BibitemOpen
  \bibfield  {author} {\bibinfo {author} {\bibfnamefont {F.}~\bibnamefont
  {Vicentini}}, \bibinfo {author} {\bibfnamefont {F.}~\bibnamefont {Minganti}},
  \bibinfo {author} {\bibfnamefont {A.}~\bibnamefont {Biella}}, \bibinfo
  {author} {\bibfnamefont {G.}~\bibnamefont {Orso}}, \ and\ \bibinfo {author}
  {\bibfnamefont {C.}~\bibnamefont {Ciuti}},\ }\href {\doibase
  10.1103/PhysRevA.99.032115} {\bibfield  {journal} {\bibinfo  {journal} {Phys.
  Rev. A}\ }\textbf {\bibinfo {volume} {99}},\ \bibinfo {pages} {032115}
  (\bibinfo {year} {2019}{\natexlab{a}})}\BibitemShut {NoStop}%
\bibitem [{\citenamefont {Sugiura}\ and\ \citenamefont
  {Shimizu}(2012)}]{sugiura2012}%
  \BibitemOpen
  \bibfield  {author} {\bibinfo {author} {\bibfnamefont {S.}~\bibnamefont
  {Sugiura}}\ and\ \bibinfo {author} {\bibfnamefont {A.}~\bibnamefont
  {Shimizu}},\ }\href {\doibase 10.1103/PhysRevLett.108.240401} {\bibfield
  {journal} {\bibinfo  {journal} {Physical Review Letters}\ }\textbf {\bibinfo
  {volume} {108}},\ \bibinfo {pages} {240401} (\bibinfo {year}
  {2012})}\BibitemShut {NoStop}%
\bibitem [{\citenamefont {Sugiura}\ and\ \citenamefont
  {Shimizu}(2013)}]{sugiura2013}%
  \BibitemOpen
  \bibfield  {author} {\bibinfo {author} {\bibfnamefont {S.}~\bibnamefont
  {Sugiura}}\ and\ \bibinfo {author} {\bibfnamefont {A.}~\bibnamefont
  {Shimizu}},\ }\href {\doibase 10.1103/PhysRevLett.111.010401} {\bibfield
  {journal} {\bibinfo  {journal} {Physical Review Letters}\ }\textbf {\bibinfo
  {volume} {111}},\ \bibinfo {pages} {010401} (\bibinfo {year}
  {2013})}\BibitemShut {NoStop}%
\bibitem [{\citenamefont {Hendry}\ \emph {et~al.}(2022)\citenamefont {Hendry},
  \citenamefont {Chen},\ and\ \citenamefont {Feiguin}}]{hendry2022neural}%
  \BibitemOpen
  \bibfield  {author} {\bibinfo {author} {\bibfnamefont {D.}~\bibnamefont
  {Hendry}}, \bibinfo {author} {\bibfnamefont {H.}~\bibnamefont {Chen}}, \ and\
  \bibinfo {author} {\bibfnamefont {A.}~\bibnamefont {Feiguin}},\ }\href
  {\doibase 10.1103/PhysRevB.106.165111} {\bibfield  {journal} {\bibinfo
  {journal} {Phys. Rev. B}\ }\textbf {\bibinfo {volume} {106}},\ \bibinfo
  {pages} {165111} (\bibinfo {year} {2022})}\BibitemShut {NoStop}%
\bibitem [{\citenamefont {Takai}\ \emph {et~al.}(2016)\citenamefont {Takai},
  \citenamefont {Ido}, \citenamefont {Misawa}, \citenamefont {Yamaji},\ and\
  \citenamefont {Imada}}]{takai2016finite}%
  \BibitemOpen
  \bibfield  {author} {\bibinfo {author} {\bibfnamefont {K.}~\bibnamefont
  {Takai}}, \bibinfo {author} {\bibfnamefont {K.}~\bibnamefont {Ido}}, \bibinfo
  {author} {\bibfnamefont {T.}~\bibnamefont {Misawa}}, \bibinfo {author}
  {\bibfnamefont {Y.}~\bibnamefont {Yamaji}}, \ and\ \bibinfo {author}
  {\bibfnamefont {M.}~\bibnamefont {Imada}},\ }\href {\doibase
  10.7566/JPSJ.85.034601} {\bibfield  {journal} {\bibinfo  {journal} {Journal
  of the Physical Society of Japan}\ }\textbf {\bibinfo {volume} {85}},\
  \bibinfo {pages} {034601} (\bibinfo {year} {2016})}\BibitemShut {NoStop}%
\bibitem [{\citenamefont {White}(2009)}]{white2009}%
  \BibitemOpen
  \bibfield  {author} {\bibinfo {author} {\bibfnamefont {S.~R.}\ \bibnamefont
  {White}},\ }\href {\doibase 10.1103/PhysRevLett.102.190601} {\bibfield
  {journal} {\bibinfo  {journal} {Physical Review Letters}\ }\textbf {\bibinfo
  {volume} {102}},\ \bibinfo {pages} {190601} (\bibinfo {year}
  {2009})}\BibitemShut {NoStop}%
\bibitem [{\citenamefont {Stoudenmire}\ and\ \citenamefont
  {White}(2010)}]{stoudenmire2010}%
  \BibitemOpen
  \bibfield  {author} {\bibinfo {author} {\bibfnamefont {E.~M.}\ \bibnamefont
  {Stoudenmire}}\ and\ \bibinfo {author} {\bibfnamefont {S.~R.}\ \bibnamefont
  {White}},\ }\href {\doibase 10.1088/1367-2630/12/5/055026} {\bibfield
  {journal} {\bibinfo  {journal} {New Journal of Physics}\ }\textbf {\bibinfo
  {volume} {12}},\ \bibinfo {pages} {055026} (\bibinfo {year}
  {2010})}\BibitemShut {NoStop}%
\bibitem [{\citenamefont {Bruognolo}\ \emph {et~al.}(2017)\citenamefont
  {Bruognolo}, \citenamefont {Zhu}, \citenamefont {White},\ and\ \citenamefont
  {Stoudenmire}}]{bruognolo2017}%
  \BibitemOpen
  \bibfield  {author} {\bibinfo {author} {\bibfnamefont {B.}~\bibnamefont
  {Bruognolo}}, \bibinfo {author} {\bibfnamefont {Z.}~\bibnamefont {Zhu}},
  \bibinfo {author} {\bibfnamefont {S.~R.}\ \bibnamefont {White}}, \ and\
  \bibinfo {author} {\bibfnamefont {E.~M.}\ \bibnamefont {Stoudenmire}},\
  }\href {\doibase 10.48550/arXiv.1705.05578} {\enquote {\bibinfo {title}
  {Matrix product state techniques for two-dimensional systems at finite
  temperature},}\ } (\bibinfo {year} {2017}),\ \Eprint
  {http://arxiv.org/abs/1705.05578} {arXiv:1705.05578 [cond-mat]} \BibitemShut
  {NoStop}%
\bibitem [{\citenamefont {Carleo}\ and\ \citenamefont
  {Troyer}(2017)}]{carleo2017}%
  \BibitemOpen
  \bibfield  {author} {\bibinfo {author} {\bibfnamefont {G.}~\bibnamefont
  {Carleo}}\ and\ \bibinfo {author} {\bibfnamefont {M.}~\bibnamefont
  {Troyer}},\ }\href {\doibase 10.1126/science.aag2302} {\bibfield  {journal}
  {\bibinfo  {journal} {Science}\ }\textbf {\bibinfo {volume} {355}},\ \bibinfo
  {pages} {602} (\bibinfo {year} {2017})}\BibitemShut {NoStop}%
\bibitem [{\citenamefont {Wu}\ \emph {et~al.}(2023)\citenamefont {Wu},
  \citenamefont {Rossi}, \citenamefont {Vicentini}, \citenamefont
  {Astrakhantsev}, \citenamefont {Becca}, \citenamefont {Cao}, \citenamefont
  {Carrasquilla}, \citenamefont {Ferrari}, \citenamefont {Georges},
  \citenamefont {Hibat-Allah} \emph {et~al.}}]{wu2023variational}%
  \BibitemOpen
  \bibfield  {author} {\bibinfo {author} {\bibfnamefont {D.}~\bibnamefont
  {Wu}}, \bibinfo {author} {\bibfnamefont {R.}~\bibnamefont {Rossi}}, \bibinfo
  {author} {\bibfnamefont {F.}~\bibnamefont {Vicentini}}, \bibinfo {author}
  {\bibfnamefont {N.}~\bibnamefont {Astrakhantsev}}, \bibinfo {author}
  {\bibfnamefont {F.}~\bibnamefont {Becca}}, \bibinfo {author} {\bibfnamefont
  {X.}~\bibnamefont {Cao}}, \bibinfo {author} {\bibfnamefont {J.}~\bibnamefont
  {Carrasquilla}}, \bibinfo {author} {\bibfnamefont {F.}~\bibnamefont
  {Ferrari}}, \bibinfo {author} {\bibfnamefont {A.}~\bibnamefont {Georges}},
  \bibinfo {author} {\bibfnamefont {M.}~\bibnamefont {Hibat-Allah}},  \emph
  {et~al.},\ }\href {\doibase 10.48550/arXiv.2302.04919} {\bibfield  {journal}
  {\bibinfo  {journal} {arXiv preprint arXiv:2302.04919}\ } (\bibinfo {year}
  {2023}),\ 10.48550/arXiv.2302.04919}\BibitemShut {NoStop}%
\bibitem [{\citenamefont {Schmitt}\ and\ \citenamefont
  {Heyl}(2020)}]{schmitt2020dynamics}%
  \BibitemOpen
  \bibfield  {author} {\bibinfo {author} {\bibfnamefont {M.}~\bibnamefont
  {Schmitt}}\ and\ \bibinfo {author} {\bibfnamefont {M.}~\bibnamefont {Heyl}},\
  }\href {\doibase 10.1103/PhysRevLett.125.100503} {\bibfield  {journal}
  {\bibinfo  {journal} {Phys. Rev. Lett.}\ }\textbf {\bibinfo {volume} {125}},\
  \bibinfo {pages} {100503} (\bibinfo {year} {2020})}\BibitemShut {NoStop}%
\bibitem [{\citenamefont {Guti{\'e}rrez}\ and\ \citenamefont
  {Mendl}(2022)}]{gutierrez2022real}%
  \BibitemOpen
  \bibfield  {author} {\bibinfo {author} {\bibfnamefont {I.~L.}\ \bibnamefont
  {Guti{\'e}rrez}}\ and\ \bibinfo {author} {\bibfnamefont {C.~B.}\ \bibnamefont
  {Mendl}},\ }\href {\doibase 10.22331/q-2022-01-20-627} {\bibfield  {journal}
  {\bibinfo  {journal} {Quantum}\ }\textbf {\bibinfo {volume} {6}},\ \bibinfo
  {pages} {627} (\bibinfo {year} {2022})}\BibitemShut {NoStop}%
\bibitem [{\citenamefont {Donatella}\ \emph {et~al.}(2023)\citenamefont
  {Donatella}, \citenamefont {Denis}, \citenamefont {Le~Boit\'e},\ and\
  \citenamefont {Ciuti}}]{donatella2022}%
  \BibitemOpen
  \bibfield  {author} {\bibinfo {author} {\bibfnamefont {K.}~\bibnamefont
  {Donatella}}, \bibinfo {author} {\bibfnamefont {Z.}~\bibnamefont {Denis}},
  \bibinfo {author} {\bibfnamefont {A.}~\bibnamefont {Le~Boit\'e}}, \ and\
  \bibinfo {author} {\bibfnamefont {C.}~\bibnamefont {Ciuti}},\ }\href
  {\doibase 10.1103/PhysRevA.108.022210} {\bibfield  {journal} {\bibinfo
  {journal} {Phys. Rev. A}\ }\textbf {\bibinfo {volume} {108}},\ \bibinfo
  {pages} {022210} (\bibinfo {year} {2023})}\BibitemShut {NoStop}%
\bibitem [{\citenamefont {Sinibaldi}\ \emph {et~al.}(2023)\citenamefont
  {Sinibaldi}, \citenamefont {Giuliani}, \citenamefont {Carleo},\ and\
  \citenamefont {Vicentini}}]{sinibaldi2023unbiasing}%
  \BibitemOpen
  \bibfield  {author} {\bibinfo {author} {\bibfnamefont {A.}~\bibnamefont
  {Sinibaldi}}, \bibinfo {author} {\bibfnamefont {C.}~\bibnamefont {Giuliani}},
  \bibinfo {author} {\bibfnamefont {G.}~\bibnamefont {Carleo}}, \ and\ \bibinfo
  {author} {\bibfnamefont {F.}~\bibnamefont {Vicentini}},\ }\href {\doibase
  10.48550/arXiv.2305.14294} {\bibfield  {journal} {\bibinfo  {journal} {arXiv
  preprint arXiv:2305.14294}\ } (\bibinfo {year} {2023}),\
  10.48550/arXiv.2305.14294}\BibitemShut {NoStop}%
\bibitem [{\citenamefont {Medvidovi{\'c}}\ and\ \citenamefont
  {Sels}(2022)}]{medvidovic2022towards}%
  \BibitemOpen
  \bibfield  {author} {\bibinfo {author} {\bibfnamefont {M.}~\bibnamefont
  {Medvidovi{\'c}}}\ and\ \bibinfo {author} {\bibfnamefont {D.}~\bibnamefont
  {Sels}},\ }\href {\doibase 10.48550/arXiv.2212.11289} {\bibfield  {journal}
  {\bibinfo  {journal} {arXiv preprint arXiv:2212.11289}\ } (\bibinfo {year}
  {2022}),\ 10.48550/arXiv.2212.11289}\BibitemShut {NoStop}%
\bibitem [{\citenamefont {Deng}\ \emph {et~al.}(2017)\citenamefont {Deng},
  \citenamefont {Li},\ and\ \citenamefont {Das~Sarma}}]{deng2017}%
  \BibitemOpen
  \bibfield  {author} {\bibinfo {author} {\bibfnamefont {D.-L.}\ \bibnamefont
  {Deng}}, \bibinfo {author} {\bibfnamefont {X.}~\bibnamefont {Li}}, \ and\
  \bibinfo {author} {\bibfnamefont {S.}~\bibnamefont {Das~Sarma}},\ }\href
  {\doibase 10.1103/PhysRevX.7.021021} {\bibfield  {journal} {\bibinfo
  {journal} {Physical Review X}\ }\textbf {\bibinfo {volume} {7}},\ \bibinfo
  {pages} {021021} (\bibinfo {year} {2017})}\BibitemShut {NoStop}%
\bibitem [{\citenamefont {Glasser}\ \emph {et~al.}(2018)\citenamefont
  {Glasser}, \citenamefont {Pancotti}, \citenamefont {August}, \citenamefont
  {Rodriguez},\ and\ \citenamefont {Cirac}}]{glasser2018}%
  \BibitemOpen
  \bibfield  {author} {\bibinfo {author} {\bibfnamefont {I.}~\bibnamefont
  {Glasser}}, \bibinfo {author} {\bibfnamefont {N.}~\bibnamefont {Pancotti}},
  \bibinfo {author} {\bibfnamefont {M.}~\bibnamefont {August}}, \bibinfo
  {author} {\bibfnamefont {I.~D.}\ \bibnamefont {Rodriguez}}, \ and\ \bibinfo
  {author} {\bibfnamefont {J.~I.}\ \bibnamefont {Cirac}},\ }\href {\doibase
  10.1103/PhysRevX.8.011006} {\bibfield  {journal} {\bibinfo  {journal}
  {Physical Review X}\ }\textbf {\bibinfo {volume} {8}},\ \bibinfo {pages}
  {011006} (\bibinfo {year} {2018})}\BibitemShut {NoStop}%
\bibitem [{\citenamefont {Levine}\ \emph {et~al.}(2019)\citenamefont {Levine},
  \citenamefont {Sharir}, \citenamefont {Cohen},\ and\ \citenamefont
  {Shashua}}]{levine2019}%
  \BibitemOpen
  \bibfield  {author} {\bibinfo {author} {\bibfnamefont {Y.}~\bibnamefont
  {Levine}}, \bibinfo {author} {\bibfnamefont {O.}~\bibnamefont {Sharir}},
  \bibinfo {author} {\bibfnamefont {N.}~\bibnamefont {Cohen}}, \ and\ \bibinfo
  {author} {\bibfnamefont {A.}~\bibnamefont {Shashua}},\ }\href {\doibase
  10.1103/PhysRevLett.122.065301} {\bibfield  {journal} {\bibinfo  {journal}
  {Physical Review Letters}\ }\textbf {\bibinfo {volume} {122}},\ \bibinfo
  {pages} {065301} (\bibinfo {year} {2019})}\BibitemShut {NoStop}%
\bibitem [{\citenamefont {Sharir}\ \emph {et~al.}(2022)\citenamefont {Sharir},
  \citenamefont {Shashua},\ and\ \citenamefont {Carleo}}]{sharir2022}%
  \BibitemOpen
  \bibfield  {author} {\bibinfo {author} {\bibfnamefont {O.}~\bibnamefont
  {Sharir}}, \bibinfo {author} {\bibfnamefont {A.}~\bibnamefont {Shashua}}, \
  and\ \bibinfo {author} {\bibfnamefont {G.}~\bibnamefont {Carleo}},\ }\href
  {\doibase 10.1103/PhysRevB.106.205136} {\bibfield  {journal} {\bibinfo
  {journal} {Physical Review B}\ }\textbf {\bibinfo {volume} {106}},\ \bibinfo
  {pages} {205136} (\bibinfo {year} {2022})}\BibitemShut {NoStop}%
\bibitem [{\citenamefont {Torlai}\ and\ \citenamefont
  {Melko}(2018)}]{torlai2018latent}%
  \BibitemOpen
  \bibfield  {author} {\bibinfo {author} {\bibfnamefont {G.}~\bibnamefont
  {Torlai}}\ and\ \bibinfo {author} {\bibfnamefont {R.~G.}\ \bibnamefont
  {Melko}},\ }\href {\doibase 10.1103/PhysRevLett.120.240503} {\bibfield
  {journal} {\bibinfo  {journal} {Phys. Rev. Lett.}\ }\textbf {\bibinfo
  {volume} {120}},\ \bibinfo {pages} {240503} (\bibinfo {year}
  {2018})}\BibitemShut {NoStop}%
\bibitem [{\citenamefont {Vicentini}\ \emph
  {et~al.}(2019{\natexlab{b}})\citenamefont {Vicentini}, \citenamefont
  {Biella}, \citenamefont {Regnault},\ and\ \citenamefont
  {Ciuti}}]{vicentini2019}%
  \BibitemOpen
  \bibfield  {author} {\bibinfo {author} {\bibfnamefont {F.}~\bibnamefont
  {Vicentini}}, \bibinfo {author} {\bibfnamefont {A.}~\bibnamefont {Biella}},
  \bibinfo {author} {\bibfnamefont {N.}~\bibnamefont {Regnault}}, \ and\
  \bibinfo {author} {\bibfnamefont {C.}~\bibnamefont {Ciuti}},\ }\href
  {\doibase 10.1103/PhysRevLett.122.250503} {\bibfield  {journal} {\bibinfo
  {journal} {Physical Review Letters}\ }\textbf {\bibinfo {volume} {122}},\
  \bibinfo {pages} {250503} (\bibinfo {year} {2019}{\natexlab{b}})}\BibitemShut
  {NoStop}%
\bibitem [{\citenamefont {Nagy}\ and\ \citenamefont {Savona}(2019)}]{nagy2019}%
  \BibitemOpen
  \bibfield  {author} {\bibinfo {author} {\bibfnamefont {A.}~\bibnamefont
  {Nagy}}\ and\ \bibinfo {author} {\bibfnamefont {V.}~\bibnamefont {Savona}},\
  }\href {\doibase 10.1103/PhysRevLett.122.250501} {\bibfield  {journal}
  {\bibinfo  {journal} {Physical Review Letters}\ }\textbf {\bibinfo {volume}
  {122}},\ \bibinfo {pages} {250501} (\bibinfo {year} {2019})}\BibitemShut
  {NoStop}%
\bibitem [{\citenamefont {Hartmann}\ and\ \citenamefont
  {Carleo}(2019)}]{hartmann2019}%
  \BibitemOpen
  \bibfield  {author} {\bibinfo {author} {\bibfnamefont {M.~J.}\ \bibnamefont
  {Hartmann}}\ and\ \bibinfo {author} {\bibfnamefont {G.}~\bibnamefont
  {Carleo}},\ }\href {\doibase 10.1103/PhysRevLett.122.250502} {\bibfield
  {journal} {\bibinfo  {journal} {Physical Review Letters}\ }\textbf {\bibinfo
  {volume} {122}},\ \bibinfo {pages} {250502} (\bibinfo {year}
  {2019})}\BibitemShut {NoStop}%
\bibitem [{\citenamefont {Yoshioka}\ and\ \citenamefont
  {Hamazaki}(2019)}]{yoshioka2019}%
  \BibitemOpen
  \bibfield  {author} {\bibinfo {author} {\bibfnamefont {N.}~\bibnamefont
  {Yoshioka}}\ and\ \bibinfo {author} {\bibfnamefont {R.}~\bibnamefont
  {Hamazaki}},\ }\href {\doibase 10.1103/PhysRevB.99.214306} {\bibfield
  {journal} {\bibinfo  {journal} {Physical Review B}\ }\textbf {\bibinfo
  {volume} {99}},\ \bibinfo {pages} {214306} (\bibinfo {year}
  {2019})}\BibitemShut {NoStop}%
\bibitem [{\citenamefont {Nomura}\ \emph {et~al.}(2021)\citenamefont {Nomura},
  \citenamefont {Yoshioka},\ and\ \citenamefont {Nori}}]{nomura2021}%
  \BibitemOpen
  \bibfield  {author} {\bibinfo {author} {\bibfnamefont {Y.}~\bibnamefont
  {Nomura}}, \bibinfo {author} {\bibfnamefont {N.}~\bibnamefont {Yoshioka}}, \
  and\ \bibinfo {author} {\bibfnamefont {F.}~\bibnamefont {Nori}},\ }\href
  {\doibase 10.1103/PhysRevLett.127.060601} {\bibfield  {journal} {\bibinfo
  {journal} {Physical Review Letters}\ }\textbf {\bibinfo {volume} {127}},\
  \bibinfo {pages} {060601} (\bibinfo {year} {2021})}\BibitemShut {NoStop}%
\bibitem [{\citenamefont {Carrasquilla}\ \emph {et~al.}(2019)\citenamefont
  {Carrasquilla}, \citenamefont {Torlai}, \citenamefont {Melko},\ and\
  \citenamefont {Aolita}}]{carrasquilla2019}%
  \BibitemOpen
  \bibfield  {author} {\bibinfo {author} {\bibfnamefont {J.}~\bibnamefont
  {Carrasquilla}}, \bibinfo {author} {\bibfnamefont {G.}~\bibnamefont
  {Torlai}}, \bibinfo {author} {\bibfnamefont {R.~G.}\ \bibnamefont {Melko}}, \
  and\ \bibinfo {author} {\bibfnamefont {L.}~\bibnamefont {Aolita}},\ }\href
  {\doibase 10.1038/s42256-019-0028-1} {\bibfield  {journal} {\bibinfo
  {journal} {Nature Machine Intelligence}\ }\textbf {\bibinfo {volume} {1}},\
  \bibinfo {pages} {155} (\bibinfo {year} {2019})}\BibitemShut {NoStop}%
\bibitem [{\citenamefont {Carrasquilla}\ \emph {et~al.}(2021)\citenamefont
  {Carrasquilla}, \citenamefont {Luo}, \citenamefont {P{\'e}rez}, \citenamefont
  {Milsted}, \citenamefont {Clark}, \citenamefont {Volkovs},\ and\
  \citenamefont {Aolita}}]{carrasquilla2021}%
  \BibitemOpen
  \bibfield  {author} {\bibinfo {author} {\bibfnamefont {J.}~\bibnamefont
  {Carrasquilla}}, \bibinfo {author} {\bibfnamefont {D.}~\bibnamefont {Luo}},
  \bibinfo {author} {\bibfnamefont {F.}~\bibnamefont {P{\'e}rez}}, \bibinfo
  {author} {\bibfnamefont {A.}~\bibnamefont {Milsted}}, \bibinfo {author}
  {\bibfnamefont {B.~K.}\ \bibnamefont {Clark}}, \bibinfo {author}
  {\bibfnamefont {M.}~\bibnamefont {Volkovs}}, \ and\ \bibinfo {author}
  {\bibfnamefont {L.}~\bibnamefont {Aolita}},\ }\href {\doibase
  10.1103/PhysRevA.104.032610} {\bibfield  {journal} {\bibinfo  {journal}
  {Physical Review A}\ }\textbf {\bibinfo {volume} {104}},\ \bibinfo {pages}
  {032610} (\bibinfo {year} {2021})}\BibitemShut {NoStop}%
\bibitem [{\citenamefont {Luo}\ \emph {et~al.}(2022)\citenamefont {Luo},
  \citenamefont {Chen}, \citenamefont {Carrasquilla},\ and\ \citenamefont
  {Clark}}]{luo2022}%
  \BibitemOpen
  \bibfield  {author} {\bibinfo {author} {\bibfnamefont {D.}~\bibnamefont
  {Luo}}, \bibinfo {author} {\bibfnamefont {Z.}~\bibnamefont {Chen}}, \bibinfo
  {author} {\bibfnamefont {J.}~\bibnamefont {Carrasquilla}}, \ and\ \bibinfo
  {author} {\bibfnamefont {B.~K.}\ \bibnamefont {Clark}},\ }\href {\doibase
  10.1103/PhysRevLett.128.090501} {\bibfield  {journal} {\bibinfo  {journal}
  {Physical Review Letters}\ }\textbf {\bibinfo {volume} {128}},\ \bibinfo
  {pages} {090501} (\bibinfo {year} {2022})}\BibitemShut {NoStop}%
\bibitem [{\citenamefont {Reh}\ \emph {et~al.}(2021)\citenamefont {Reh},
  \citenamefont {Schmitt},\ and\ \citenamefont {G\"arttner}}]{reh2021time}%
  \BibitemOpen
  \bibfield  {author} {\bibinfo {author} {\bibfnamefont {M.}~\bibnamefont
  {Reh}}, \bibinfo {author} {\bibfnamefont {M.}~\bibnamefont {Schmitt}}, \ and\
  \bibinfo {author} {\bibfnamefont {M.}~\bibnamefont {G\"arttner}},\ }\href
  {\doibase 10.1103/PhysRevLett.127.230501} {\bibfield  {journal} {\bibinfo
  {journal} {Phys. Rev. Lett.}\ }\textbf {\bibinfo {volume} {127}},\ \bibinfo
  {pages} {230501} (\bibinfo {year} {2021})}\BibitemShut {NoStop}%
\bibitem [{\citenamefont {Vicentini}\ \emph
  {et~al.}(2022{\natexlab{a}})\citenamefont {Vicentini}, \citenamefont
  {Rossi},\ and\ \citenamefont {Carleo}}]{vicentini2022}%
  \BibitemOpen
  \bibfield  {author} {\bibinfo {author} {\bibfnamefont {F.}~\bibnamefont
  {Vicentini}}, \bibinfo {author} {\bibfnamefont {R.}~\bibnamefont {Rossi}}, \
  and\ \bibinfo {author} {\bibfnamefont {G.}~\bibnamefont {Carleo}},\ }\href
  {\doibase 10.48550/arXiv.2206.13488} {\enquote {\bibinfo {title}
  {Positive-definite parametrization of mixed quantum states with deep neural
  networks},}\ } (\bibinfo {year} {2022}{\natexlab{a}}),\ \Eprint
  {http://arxiv.org/abs/2206.13488} {arXiv:2206.13488} \BibitemShut {NoStop}%
\bibitem [{\citenamefont {Takahashi}\ and\ \citenamefont
  {Umezawa}(1975)}]{takahashi1975}%
  \BibitemOpen
  \bibfield  {author} {\bibinfo {author} {\bibfnamefont {Y.}~\bibnamefont
  {Takahashi}}\ and\ \bibinfo {author} {\bibfnamefont {H.}~\bibnamefont
  {Umezawa}},\ }\href@noop {} {\bibfield  {journal} {\bibinfo  {journal}
  {Collective phenomena}\ }\textbf {\bibinfo {volume} {2}},\ \bibinfo {pages}
  {55} (\bibinfo {year} {1975})}\BibitemShut {NoStop}%
\bibitem [{\citenamefont {Suzuki}(1985)}]{suzuki1985}%
  \BibitemOpen
  \bibfield  {author} {\bibinfo {author} {\bibfnamefont {M.}~\bibnamefont
  {Suzuki}},\ }\href {\doibase 10.1143/JPSJ.54.4483} {\bibfield  {journal}
  {\bibinfo  {journal} {Journal of the Physical Society of Japan}\ }\textbf
  {\bibinfo {volume} {54}},\ \bibinfo {pages} {4483} (\bibinfo {year}
  {1985})}\BibitemShut {NoStop}%
\bibitem [{\citenamefont {Israel}(1976)}]{israel1976}%
  \BibitemOpen
  \bibfield  {author} {\bibinfo {author} {\bibfnamefont {W.}~\bibnamefont
  {Israel}},\ }\href {\doibase 10.1016/0375-9601(76)90178-X} {\bibfield
  {journal} {\bibinfo  {journal} {Physics Letters A}\ }\textbf {\bibinfo
  {volume} {57}},\ \bibinfo {pages} {107} (\bibinfo {year} {1976})}\BibitemShut
  {NoStop}%
\bibitem [{\citenamefont {Harsha}\ \emph
  {et~al.}(2019{\natexlab{a}})\citenamefont {Harsha}, \citenamefont
  {Henderson},\ and\ \citenamefont {Scuseria}}]{harsha2019}%
  \BibitemOpen
  \bibfield  {author} {\bibinfo {author} {\bibfnamefont {G.}~\bibnamefont
  {Harsha}}, \bibinfo {author} {\bibfnamefont {T.~M.}\ \bibnamefont
  {Henderson}}, \ and\ \bibinfo {author} {\bibfnamefont {G.~E.}\ \bibnamefont
  {Scuseria}},\ }\href {\doibase 10.1063/1.5089560} {\bibfield  {journal}
  {\bibinfo  {journal} {The Journal of Chemical Physics}\ }\textbf {\bibinfo
  {volume} {150}},\ \bibinfo {pages} {154109} (\bibinfo {year}
  {2019}{\natexlab{a}})}\BibitemShut {NoStop}%
\bibitem [{\citenamefont {Shushkov}\ and\ \citenamefont
  {Miller}(2019)}]{shushkov2019}%
  \BibitemOpen
  \bibfield  {author} {\bibinfo {author} {\bibfnamefont {P.}~\bibnamefont
  {Shushkov}}\ and\ \bibinfo {author} {\bibfnamefont {T.~F.}\ \bibnamefont
  {Miller}},\ }\href {\doibase 10.1063/1.5121749} {\bibfield  {journal}
  {\bibinfo  {journal} {The Journal of Chemical Physics}\ }\textbf {\bibinfo
  {volume} {151}},\ \bibinfo {pages} {134107} (\bibinfo {year}
  {2019})}\BibitemShut {NoStop}%
\bibitem [{\citenamefont {Harsha}\ \emph
  {et~al.}(2019{\natexlab{b}})\citenamefont {Harsha}, \citenamefont
  {Henderson},\ and\ \citenamefont {Scuseria}}]{harsha2019b}%
  \BibitemOpen
  \bibfield  {author} {\bibinfo {author} {\bibfnamefont {G.}~\bibnamefont
  {Harsha}}, \bibinfo {author} {\bibfnamefont {T.~M.}\ \bibnamefont
  {Henderson}}, \ and\ \bibinfo {author} {\bibfnamefont {G.~E.}\ \bibnamefont
  {Scuseria}},\ }\href {\doibase 10.1021/acs.jctc.9b00744} {\bibfield
  {journal} {\bibinfo  {journal} {Journal of Chemical Theory and Computation}\
  }\textbf {\bibinfo {volume} {15}},\ \bibinfo {pages} {6127} (\bibinfo {year}
  {2019}{\natexlab{b}})}\BibitemShut {NoStop}%
\bibitem [{\citenamefont {Harsha}\ \emph {et~al.}(2020)\citenamefont {Harsha},
  \citenamefont {Henderson},\ and\ \citenamefont {Scuseria}}]{harsha2020a}%
  \BibitemOpen
  \bibfield  {author} {\bibinfo {author} {\bibfnamefont {G.}~\bibnamefont
  {Harsha}}, \bibinfo {author} {\bibfnamefont {T.~M.}\ \bibnamefont
  {Henderson}}, \ and\ \bibinfo {author} {\bibfnamefont {G.~E.}\ \bibnamefont
  {Scuseria}},\ }\href {\doibase 10.1063/5.0022702} {\bibfield  {journal}
  {\bibinfo  {journal} {The Journal of Chemical Physics}\ }\textbf {\bibinfo
  {volume} {153}},\ \bibinfo {pages} {124115} (\bibinfo {year}
  {2020})}\BibitemShut {NoStop}%
\bibitem [{\citenamefont {Chapman}\ \emph {et~al.}(2019)\citenamefont
  {Chapman}, \citenamefont {Eisert}, \citenamefont {Hackl}, \citenamefont
  {Heller}, \citenamefont {Jefferson}, \citenamefont {Marrochio},\ and\
  \citenamefont {Myers}}]{chapman2019}%
  \BibitemOpen
  \bibfield  {author} {\bibinfo {author} {\bibfnamefont {S.}~\bibnamefont
  {Chapman}}, \bibinfo {author} {\bibfnamefont {J.}~\bibnamefont {Eisert}},
  \bibinfo {author} {\bibfnamefont {L.}~\bibnamefont {Hackl}}, \bibinfo
  {author} {\bibfnamefont {M.~P.}\ \bibnamefont {Heller}}, \bibinfo {author}
  {\bibfnamefont {R.}~\bibnamefont {Jefferson}}, \bibinfo {author}
  {\bibfnamefont {H.}~\bibnamefont {Marrochio}}, \ and\ \bibinfo {author}
  {\bibfnamefont {R.}~\bibnamefont {Myers}},\ }\href {\doibase
  10.21468/SciPostPhys.6.3.034} {\bibfield  {journal} {\bibinfo  {journal}
  {SciPost Physics}\ }\textbf {\bibinfo {volume} {6}},\ \bibinfo {pages} {034}
  (\bibinfo {year} {2019})}\BibitemShut {NoStop}%
\bibitem [{\citenamefont {Lee}\ \emph {et~al.}(2022)\citenamefont {Lee},
  \citenamefont {Zhang}, \citenamefont {Hsieh}, \citenamefont {Zhang},\ and\
  \citenamefont {Shi}}]{lee2022variational}%
  \BibitemOpen
  \bibfield  {author} {\bibinfo {author} {\bibfnamefont {C.~K.}\ \bibnamefont
  {Lee}}, \bibinfo {author} {\bibfnamefont {S.-X.}\ \bibnamefont {Zhang}},
  \bibinfo {author} {\bibfnamefont {C.-Y.}\ \bibnamefont {Hsieh}}, \bibinfo
  {author} {\bibfnamefont {S.}~\bibnamefont {Zhang}}, \ and\ \bibinfo {author}
  {\bibfnamefont {L.}~\bibnamefont {Shi}},\ }\href {\doibase
  10.48550/arXiv.2206.05571} {\bibfield  {journal} {\bibinfo  {journal} {arXiv
  preprint arXiv:2206.05571}\ } (\bibinfo {year} {2022}),\
  10.48550/arXiv.2206.05571}\BibitemShut {NoStop}%
\bibitem [{\citenamefont {Zhu}\ \emph {et~al.}(2020)\citenamefont {Zhu},
  \citenamefont {Johri}, \citenamefont {Linke}, \citenamefont {Landsman},
  \citenamefont {Alderete}, \citenamefont {Nguyen}, \citenamefont {Matsuura},
  \citenamefont {Hsieh},\ and\ \citenamefont {Monroe}}]{zhu2020generation}%
  \BibitemOpen
  \bibfield  {author} {\bibinfo {author} {\bibfnamefont {D.}~\bibnamefont
  {Zhu}}, \bibinfo {author} {\bibfnamefont {S.}~\bibnamefont {Johri}}, \bibinfo
  {author} {\bibfnamefont {N.~M.}\ \bibnamefont {Linke}}, \bibinfo {author}
  {\bibfnamefont {K.~A.}\ \bibnamefont {Landsman}}, \bibinfo {author}
  {\bibfnamefont {C.~H.}\ \bibnamefont {Alderete}}, \bibinfo {author}
  {\bibfnamefont {N.~H.}\ \bibnamefont {Nguyen}}, \bibinfo {author}
  {\bibfnamefont {A.~Y.}\ \bibnamefont {Matsuura}}, \bibinfo {author}
  {\bibfnamefont {T.~H.}\ \bibnamefont {Hsieh}}, \ and\ \bibinfo {author}
  {\bibfnamefont {C.}~\bibnamefont {Monroe}},\ }\href {\doibase
  10.1073/pnas.2006337117} {\bibfield  {journal} {\bibinfo  {journal}
  {Proceedings of the National Academy of Sciences}\ }\textbf {\bibinfo
  {volume} {117}},\ \bibinfo {pages} {25402} (\bibinfo {year}
  {2020})}\BibitemShut {NoStop}%
\bibitem [{\citenamefont {Gyamfi}(2020)}]{gyamfi2020}%
  \BibitemOpen
  \bibfield  {author} {\bibinfo {author} {\bibfnamefont {J.~A.}\ \bibnamefont
  {Gyamfi}},\ }\href {\doibase 10.1088/1361-6404/ab9fdd} {\bibfield  {journal}
  {\bibinfo  {journal} {European Journal of Physics}\ }\textbf {\bibinfo
  {volume} {41}},\ \bibinfo {pages} {063002} (\bibinfo {year}
  {2020})}\BibitemShut {NoStop}%
\bibitem [{\citenamefont {Arimitsu}\ and\ \citenamefont
  {Umezawa}(1985)}]{arimitsu1985}%
  \BibitemOpen
  \bibfield  {author} {\bibinfo {author} {\bibfnamefont {T.}~\bibnamefont
  {Arimitsu}}\ and\ \bibinfo {author} {\bibfnamefont {H.}~\bibnamefont
  {Umezawa}},\ }\href {\doibase 10.1143/PTP.74.429} {\bibfield  {journal}
  {\bibinfo  {journal} {Progress of Theoretical Physics}\ }\textbf {\bibinfo
  {volume} {74}},\ \bibinfo {pages} {429} (\bibinfo {year} {1985})}\BibitemShut
  {NoStop}%
\bibitem [{Note1()}]{Note1}%
  \BibitemOpen
  \bibinfo {note} {Supplemental Material, including references~\cite
  {kothe2023liouville, rackauckas2017differentialequations}.}\BibitemShut
  {Stop}%
\bibitem [{\citenamefont {Sharir}\ \emph {et~al.}(2020)\citenamefont {Sharir},
  \citenamefont {Levine}, \citenamefont {Wies}, \citenamefont {Carleo},\ and\
  \citenamefont {Shashua}}]{sharir2020}%
  \BibitemOpen
  \bibfield  {author} {\bibinfo {author} {\bibfnamefont {O.}~\bibnamefont
  {Sharir}}, \bibinfo {author} {\bibfnamefont {Y.}~\bibnamefont {Levine}},
  \bibinfo {author} {\bibfnamefont {N.}~\bibnamefont {Wies}}, \bibinfo {author}
  {\bibfnamefont {G.}~\bibnamefont {Carleo}}, \ and\ \bibinfo {author}
  {\bibfnamefont {A.}~\bibnamefont {Shashua}},\ }\href {\doibase
  10.1103/PhysRevLett.124.020503} {\bibfield  {journal} {\bibinfo  {journal}
  {Physical Review Letters}\ }\textbf {\bibinfo {volume} {124}},\ \bibinfo
  {pages} {020503} (\bibinfo {year} {2020})}\BibitemShut {NoStop}%
\bibitem [{\citenamefont {{Hibat-Allah}}\ \emph {et~al.}(2020)\citenamefont
  {{Hibat-Allah}}, \citenamefont {Ganahl}, \citenamefont {Hayward},
  \citenamefont {Melko},\ and\ \citenamefont {Carrasquilla}}]{hibat-allah2020}%
  \BibitemOpen
  \bibfield  {author} {\bibinfo {author} {\bibfnamefont {M.}~\bibnamefont
  {{Hibat-Allah}}}, \bibinfo {author} {\bibfnamefont {M.}~\bibnamefont
  {Ganahl}}, \bibinfo {author} {\bibfnamefont {L.~E.}\ \bibnamefont {Hayward}},
  \bibinfo {author} {\bibfnamefont {R.~G.}\ \bibnamefont {Melko}}, \ and\
  \bibinfo {author} {\bibfnamefont {J.}~\bibnamefont {Carrasquilla}},\ }\href
  {\doibase 10.1103/PhysRevResearch.2.023358} {\bibfield  {journal} {\bibinfo
  {journal} {Physical Review Research}\ }\textbf {\bibinfo {volume} {2}},\
  \bibinfo {pages} {023358} (\bibinfo {year} {2020})}\BibitemShut {NoStop}%
\bibitem [{\citenamefont {Wu}\ \emph {et~al.}(2022)\citenamefont {Wu},
  \citenamefont {Rossi}, \citenamefont {Vicentini},\ and\ \citenamefont
  {Carleo}}]{wu2022}%
  \BibitemOpen
  \bibfield  {author} {\bibinfo {author} {\bibfnamefont {D.}~\bibnamefont
  {Wu}}, \bibinfo {author} {\bibfnamefont {R.}~\bibnamefont {Rossi}}, \bibinfo
  {author} {\bibfnamefont {F.}~\bibnamefont {Vicentini}}, \ and\ \bibinfo
  {author} {\bibfnamefont {G.}~\bibnamefont {Carleo}},\ }\href {\doibase
  10.48550/arXiv.2206.12363} {\enquote {\bibinfo {title} {From {{Tensor Network
  Quantum States}} to {{Tensorial Recurrent Neural Networks}}},}\ } (\bibinfo
  {year} {2022}),\ \Eprint {http://arxiv.org/abs/2206.12363} {arXiv:2206.12363}
  \BibitemShut {NoStop}%
\bibitem [{\citenamefont {Hibat-Allah}\ \emph {et~al.}(2023)\citenamefont
  {Hibat-Allah}, \citenamefont {Melko},\ and\ \citenamefont
  {Carrasquilla}}]{hibat2023investigating}%
  \BibitemOpen
  \bibfield  {author} {\bibinfo {author} {\bibfnamefont {M.}~\bibnamefont
  {Hibat-Allah}}, \bibinfo {author} {\bibfnamefont {R.~G.}\ \bibnamefont
  {Melko}}, \ and\ \bibinfo {author} {\bibfnamefont {J.}~\bibnamefont
  {Carrasquilla}},\ }\href {\doibase 10.1103/physrevb.108.075152} {\bibfield
  {journal} {\bibinfo  {journal} {Physical Review B}\ }\textbf {\bibinfo
  {volume} {108}} (\bibinfo {year} {2023}),\
  10.1103/physrevb.108.075152}\BibitemShut {NoStop}%
\bibitem [{\citenamefont {Yuan}\ \emph {et~al.}(2019)\citenamefont {Yuan},
  \citenamefont {Endo}, \citenamefont {Zhao}, \citenamefont {Li},\ and\
  \citenamefont {Benjamin}}]{yuan2019}%
  \BibitemOpen
  \bibfield  {author} {\bibinfo {author} {\bibfnamefont {X.}~\bibnamefont
  {Yuan}}, \bibinfo {author} {\bibfnamefont {S.}~\bibnamefont {Endo}}, \bibinfo
  {author} {\bibfnamefont {Q.}~\bibnamefont {Zhao}}, \bibinfo {author}
  {\bibfnamefont {Y.}~\bibnamefont {Li}}, \ and\ \bibinfo {author}
  {\bibfnamefont {S.~C.}\ \bibnamefont {Benjamin}},\ }\href {\doibase
  10.22331/q-2019-10-07-191} {\bibfield  {journal} {\bibinfo  {journal}
  {Quantum}\ }\textbf {\bibinfo {volume} {3}},\ \bibinfo {pages} {191}
  (\bibinfo {year} {2019})}\BibitemShut {NoStop}%
\bibitem [{\citenamefont {Broeckhove}\ \emph {et~al.}(1988)\citenamefont
  {Broeckhove}, \citenamefont {Lathouwers}, \citenamefont {Kesteloot},\ and\
  \citenamefont {{Van Leuven}}}]{broeckhove1988equivalence}%
  \BibitemOpen
  \bibfield  {author} {\bibinfo {author} {\bibfnamefont {J.}~\bibnamefont
  {Broeckhove}}, \bibinfo {author} {\bibfnamefont {L.}~\bibnamefont
  {Lathouwers}}, \bibinfo {author} {\bibfnamefont {E.}~\bibnamefont
  {Kesteloot}}, \ and\ \bibinfo {author} {\bibfnamefont {P.}~\bibnamefont {{Van
  Leuven}}},\ }\href {\doibase https://doi.org/10.1016/0009-2614(88)80380-4}
  {\bibfield  {journal} {\bibinfo  {journal} {Chemical Physics Letters}\
  }\textbf {\bibinfo {volume} {149}},\ \bibinfo {pages} {547} (\bibinfo {year}
  {1988})}\BibitemShut {NoStop}%
\bibitem [{\citenamefont {Sorella}(1998)}]{sorella1998green}%
  \BibitemOpen
  \bibfield  {author} {\bibinfo {author} {\bibfnamefont {S.}~\bibnamefont
  {Sorella}},\ }\href {\doibase 10.1103/PhysRevLett.80.4558} {\bibfield
  {journal} {\bibinfo  {journal} {Phys. Rev. Lett.}\ }\textbf {\bibinfo
  {volume} {80}},\ \bibinfo {pages} {4558} (\bibinfo {year}
  {1998})}\BibitemShut {NoStop}%
\bibitem [{\citenamefont {Vicentini}\ \emph
  {et~al.}(2022{\natexlab{b}})\citenamefont {Vicentini}, \citenamefont
  {Hofmann}, \citenamefont {Szab{\'o}}, \citenamefont {Wu}, \citenamefont
  {Roth}, \citenamefont {Giuliani}, \citenamefont {Pescia}, \citenamefont
  {Nys}, \citenamefont {Vargas-Calder{\'o}n}, \citenamefont {Astrakhantsev},\
  and\ \citenamefont {Carleo}}]{vicentini2022netket}%
  \BibitemOpen
  \bibfield  {author} {\bibinfo {author} {\bibfnamefont {F.}~\bibnamefont
  {Vicentini}}, \bibinfo {author} {\bibfnamefont {D.}~\bibnamefont {Hofmann}},
  \bibinfo {author} {\bibfnamefont {A.}~\bibnamefont {Szab{\'o}}}, \bibinfo
  {author} {\bibfnamefont {D.}~\bibnamefont {Wu}}, \bibinfo {author}
  {\bibfnamefont {C.}~\bibnamefont {Roth}}, \bibinfo {author} {\bibfnamefont
  {C.}~\bibnamefont {Giuliani}}, \bibinfo {author} {\bibfnamefont
  {G.}~\bibnamefont {Pescia}}, \bibinfo {author} {\bibfnamefont
  {J.}~\bibnamefont {Nys}}, \bibinfo {author} {\bibfnamefont {V.}~\bibnamefont
  {Vargas-Calder{\'o}n}}, \bibinfo {author} {\bibfnamefont {N.}~\bibnamefont
  {Astrakhantsev}}, \ and\ \bibinfo {author} {\bibfnamefont {G.}~\bibnamefont
  {Carleo}},\ }\href {\doibase 10.21468/scipostphyscodeb.7} {\bibfield
  {journal} {\bibinfo  {journal} {SciPost Physics Codebases}\ ,\ \bibinfo
  {pages} {007}} (\bibinfo {year} {2022}{\natexlab{b}})}\BibitemShut {NoStop}%
\bibitem [{\citenamefont {Carleo}\ \emph {et~al.}(2012)\citenamefont {Carleo},
  \citenamefont {Becca}, \citenamefont {Schiro},\ and\ \citenamefont
  {Fabrizio}}]{carleo_localization_2012}%
  \BibitemOpen
  \bibfield  {author} {\bibinfo {author} {\bibfnamefont {G.}~\bibnamefont
  {Carleo}}, \bibinfo {author} {\bibfnamefont {F.}~\bibnamefont {Becca}},
  \bibinfo {author} {\bibfnamefont {M.}~\bibnamefont {Schiro}}, \ and\ \bibinfo
  {author} {\bibfnamefont {M.}~\bibnamefont {Fabrizio}},\ }\href {\doibase
  10.1038/srep00243} {\bibfield  {journal} {\bibinfo  {journal} {Scientific
  Reports}\ }\textbf {\bibinfo {volume} {2}},\ \bibinfo {pages} {243} (\bibinfo
  {year} {2012})}\BibitemShut {NoStop}%
\bibitem [{\citenamefont {Carleo}\ \emph {et~al.}(2017)\citenamefont {Carleo},
  \citenamefont {Cevolani}, \citenamefont {Sanchez-Palencia},\ and\
  \citenamefont {Holzmann}}]{carleo2017unitary}%
  \BibitemOpen
  \bibfield  {author} {\bibinfo {author} {\bibfnamefont {G.}~\bibnamefont
  {Carleo}}, \bibinfo {author} {\bibfnamefont {L.}~\bibnamefont {Cevolani}},
  \bibinfo {author} {\bibfnamefont {L.}~\bibnamefont {Sanchez-Palencia}}, \
  and\ \bibinfo {author} {\bibfnamefont {M.}~\bibnamefont {Holzmann}},\ }\href
  {\doibase 10.1103/PhysRevX.7.031026} {\bibfield  {journal} {\bibinfo
  {journal} {Phys. Rev. X}\ }\textbf {\bibinfo {volume} {7}},\ \bibinfo {pages}
  {031026} (\bibinfo {year} {2017})}\BibitemShut {NoStop}%
\bibitem [{\citenamefont {Nys}\ \emph {et~al.}(2024)\citenamefont {Nys},
  \citenamefont {Pescia},\ and\ \citenamefont {Carleo}}]{nys2024ab}%
  \BibitemOpen
  \bibfield  {author} {\bibinfo {author} {\bibfnamefont {J.}~\bibnamefont
  {Nys}}, \bibinfo {author} {\bibfnamefont {G.}~\bibnamefont {Pescia}}, \ and\
  \bibinfo {author} {\bibfnamefont {G.}~\bibnamefont {Carleo}},\ }\href@noop {}
  {\enquote {\bibinfo {title} {Ab-initio variational wave functions for the
  time-dependent many-electron schr\"odinger equation},}\ } (\bibinfo {year}
  {2024}),\ \Eprint {http://arxiv.org/abs/2403.07447} {arXiv:2403.07447
  [cond-mat.str-el]} \BibitemShut {NoStop}%
\bibitem [{\citenamefont {Bl\"ote}\ and\ \citenamefont
  {Deng}(2002)}]{bloete2002}%
  \BibitemOpen
  \bibfield  {author} {\bibinfo {author} {\bibfnamefont {H.~W.~J.}\
  \bibnamefont {Bl\"ote}}\ and\ \bibinfo {author} {\bibfnamefont
  {Y.}~\bibnamefont {Deng}},\ }\href {\doibase 10.1103/PhysRevE.66.066110}
  {\bibfield  {journal} {\bibinfo  {journal} {Phys. Rev. E}\ }\textbf {\bibinfo
  {volume} {66}},\ \bibinfo {pages} {066110} (\bibinfo {year}
  {2002})}\BibitemShut {NoStop}%
\bibitem [{Note2()}]{Note2}%
  \BibitemOpen
  \bibinfo {note} {Notice also that mapping the real-time dynamics of a quantum
  system in the METTS/TPQ formalism is challenging with variational
  methods~\cite {hendry2022neural} (since an ensemble of time-evolving states
  must be traced) for small and intermediate system sizes within reach of
  numerical simulations, where many samples must be propagated.}\BibitemShut
  {Stop}%
\bibitem [{\citenamefont {Stinchcombe}(1973)}]{stinchcombe1973}%
  \BibitemOpen
  \bibfield  {author} {\bibinfo {author} {\bibfnamefont {R.~B.}\ \bibnamefont
  {Stinchcombe}},\ }\href {\doibase 10.1088/0022-3719/6/15/009} {\bibfield
  {journal} {\bibinfo  {journal} {Journal of Physics C: Solid State Physics}\
  }\textbf {\bibinfo {volume} {6}},\ \bibinfo {pages} {2459} (\bibinfo {year}
  {1973})}\BibitemShut {NoStop}%
\bibitem [{\citenamefont {Puebla}\ \emph {et~al.}(2019)\citenamefont {Puebla},
  \citenamefont {Marty},\ and\ \citenamefont {Plenio}}]{puebla2019}%
  \BibitemOpen
  \bibfield  {author} {\bibinfo {author} {\bibfnamefont {R.}~\bibnamefont
  {Puebla}}, \bibinfo {author} {\bibfnamefont {O.}~\bibnamefont {Marty}}, \
  and\ \bibinfo {author} {\bibfnamefont {M.~B.}\ \bibnamefont {Plenio}},\
  }\href {\doibase 10.1103/PhysRevA.100.032115} {\bibfield  {journal} {\bibinfo
   {journal} {Phys. Rev. A}\ }\textbf {\bibinfo {volume} {100}},\ \bibinfo
  {pages} {032115} (\bibinfo {year} {2019})}\BibitemShut {NoStop}%
\bibitem [{\citenamefont {Lin}\ and\ \citenamefont
  {Pollmann}(2022)}]{lin2022scaling}%
  \BibitemOpen
  \bibfield  {author} {\bibinfo {author} {\bibfnamefont {S.-H.}\ \bibnamefont
  {Lin}}\ and\ \bibinfo {author} {\bibfnamefont {F.}~\bibnamefont {Pollmann}},\
  }\href {\doibase 10.1002/pssb.202100172} {\bibfield  {journal} {\bibinfo
  {journal} {Physica Status Solidi (b)}\ }\textbf {\bibinfo {volume} {259}},\
  \bibinfo {pages} {2100172} (\bibinfo {year} {2022})}\BibitemShut {NoStop}%
\bibitem [{\citenamefont {Nys}\ and\ \citenamefont
  {Carleo}(2022)}]{nys2022variational}%
  \BibitemOpen
  \bibfield  {author} {\bibinfo {author} {\bibfnamefont {J.}~\bibnamefont
  {Nys}}\ and\ \bibinfo {author} {\bibfnamefont {G.}~\bibnamefont {Carleo}},\
  }\href {\doibase 10.22331/q-2022-10-13-833} {\bibfield  {journal} {\bibinfo
  {journal} {Quantum}\ }\textbf {\bibinfo {volume} {6}},\ \bibinfo {pages}
  {833} (\bibinfo {year} {2022})}\BibitemShut {NoStop}%
\bibitem [{\citenamefont {Bradbury}\ \emph {et~al.}(2018)\citenamefont
  {Bradbury}, \citenamefont {Frostig}, \citenamefont {Hawkins}, \citenamefont
  {Johnson}, \citenamefont {Leary}, \citenamefont {Maclaurin}, \citenamefont
  {Necula}, \citenamefont {Paszke}, \citenamefont {Vander{P}las}, \citenamefont
  {Wanderman-{M}ilne},\ and\ \citenamefont {Zhang}}]{jax2018github}%
  \BibitemOpen
  \bibfield  {author} {\bibinfo {author} {\bibfnamefont {J.}~\bibnamefont
  {Bradbury}}, \bibinfo {author} {\bibfnamefont {R.}~\bibnamefont {Frostig}},
  \bibinfo {author} {\bibfnamefont {P.}~\bibnamefont {Hawkins}}, \bibinfo
  {author} {\bibfnamefont {M.~J.}\ \bibnamefont {Johnson}}, \bibinfo {author}
  {\bibfnamefont {C.}~\bibnamefont {Leary}}, \bibinfo {author} {\bibfnamefont
  {D.}~\bibnamefont {Maclaurin}}, \bibinfo {author} {\bibfnamefont
  {G.}~\bibnamefont {Necula}}, \bibinfo {author} {\bibfnamefont
  {A.}~\bibnamefont {Paszke}}, \bibinfo {author} {\bibfnamefont
  {J.}~\bibnamefont {Vander{P}las}}, \bibinfo {author} {\bibfnamefont
  {S.}~\bibnamefont {Wanderman-{M}ilne}}, \ and\ \bibinfo {author}
  {\bibfnamefont {Q.}~\bibnamefont {Zhang}},\ }\href
  {http://github.com/google/jax} {\enquote {\bibinfo {title} {{JAX}: composable
  transformations of {P}ython+{N}um{P}y programs},}\ } (\bibinfo {year}
  {2018})\BibitemShut {NoStop}%
\bibitem [{\citenamefont {Häfner}\ and\ \citenamefont
  {Vicentini}(2021)}]{hafner2021mpi4jax}%
  \BibitemOpen
  \bibfield  {author} {\bibinfo {author} {\bibfnamefont {D.}~\bibnamefont
  {Häfner}}\ and\ \bibinfo {author} {\bibfnamefont {F.}~\bibnamefont
  {Vicentini}},\ }\href {\doibase 10.21105/joss.03419} {\bibfield  {journal}
  {\bibinfo  {journal} {Journal of Open Source Software}\ }\textbf {\bibinfo
  {volume} {6}},\ \bibinfo {pages} {3419} (\bibinfo {year} {2021})}\BibitemShut
  {NoStop}%
\bibitem [{\citenamefont {Kothe}\ and\ \citenamefont
  {Kirton}(2023)}]{kothe2023liouville}%
  \BibitemOpen
  \bibfield  {author} {\bibinfo {author} {\bibfnamefont {S.}~\bibnamefont
  {Kothe}}\ and\ \bibinfo {author} {\bibfnamefont {P.}~\bibnamefont {Kirton}},\
  }\href {\doibase 10.48550/arXiv.2305.13992} {\bibfield  {journal} {\bibinfo
  {journal} {arXiv preprint arXiv:2305.13992}\ } (\bibinfo {year} {2023}),\
  10.48550/arXiv.2305.13992}\BibitemShut {NoStop}%
\bibitem [{\citenamefont {Rackauckas}\ and\ \citenamefont
  {Nie}(2017)}]{rackauckas2017differentialequations}%
  \BibitemOpen
  \bibfield  {author} {\bibinfo {author} {\bibfnamefont {C.}~\bibnamefont
  {Rackauckas}}\ and\ \bibinfo {author} {\bibfnamefont {Q.}~\bibnamefont
  {Nie}},\ }\href {\doibase 10.5334/jors.151} {\bibfield  {journal} {\bibinfo
  {journal} {Journal of Open Research Software}\ }\textbf {\bibinfo {volume}
  {5}},\ \bibinfo {pages} {15} (\bibinfo {year} {2017})}\BibitemShut {NoStop}%
\bibitem [{\citenamefont {Bakshi}\ \emph {et~al.}(2024)\citenamefont {Bakshi},
  \citenamefont {Liu}, \citenamefont {Moitra},\ and\ \citenamefont
  {Tang}}]{bakshi2024high}%
  \BibitemOpen
  \bibfield  {author} {\bibinfo {author} {\bibfnamefont {A.}~\bibnamefont
  {Bakshi}}, \bibinfo {author} {\bibfnamefont {A.}~\bibnamefont {Liu}},
  \bibinfo {author} {\bibfnamefont {A.}~\bibnamefont {Moitra}}, \ and\ \bibinfo
  {author} {\bibfnamefont {E.}~\bibnamefont {Tang}},\ }\href@noop {} {\enquote
  {\bibinfo {title} {High-temperature gibbs states are unentangled and
  efficiently preparable},}\ } (\bibinfo {year} {2024}),\ \Eprint
  {http://arxiv.org/abs/2403.16850} {arXiv:2403.16850 [quant-ph]} \BibitemShut
  {NoStop}%
\bibitem [{\citenamefont {Petronilo}\ \emph {et~al.}(2021)\citenamefont
  {Petronilo}, \citenamefont {Ara{\'u}jo},\ and\ \citenamefont
  {Cruz}}]{petronilo2021simulating}%
  \BibitemOpen
  \bibfield  {author} {\bibinfo {author} {\bibfnamefont {G.}~\bibnamefont
  {Petronilo}}, \bibinfo {author} {\bibfnamefont {M.}~\bibnamefont
  {Ara{\'u}jo}}, \ and\ \bibinfo {author} {\bibfnamefont {C.}~\bibnamefont
  {Cruz}},\ }\href {\doibase 10.48550/arXiv.2111.09969} {\bibfield  {journal}
  {\bibinfo  {journal} {arXiv preprint arXiv:2111.09969}\ } (\bibinfo {year}
  {2021}),\ 10.48550/arXiv.2111.09969}\BibitemShut {NoStop}%
\end{thebibliography}%

\newpage

\appendix

\section{Computing thermal observables}\label{sec:thermal_observables}
For any arbitrary sparse Hermitian operator in the physical space $\hat{O}$ (e.g. $\hat{O} = \hat{Y}_i \otimes \hat{Y}_j$), thermal observables can be obtained through the following Monte Carlo estimator 
\begin{align}
    \expval{\smash{\hat{O}}}_\beta &= \expval{\smash{\hat{O} \otimes \tilde{1}}}_\beta \nonumber \\
    &= \mathbb{E}_{(\vb*{\sigma},\tilde{\vb*{s}}) \sim \abs{\psi_{\vb*{\theta}(\beta, t)}}^2} \left[ \sum_{\vb*{\sigma}'} \mel{\vb*{\sigma}}{\hat{O}}{\vb*{\sigma}'} \frac{\psi_{\vb*{\theta}(\beta, t)}(\vb*{\sigma}',\tilde{\vb*{s}})}{\psi_{\vb*{\theta}(\beta, t)}(\vb*{\sigma},\tilde{\vb*{s}})}\right], \label{eq:physical_Oloc}
\end{align}
where $\psi_{\vb*{\theta}(\beta, t)}$ is any of the network representations introduced in our work. For the case of ARNNO, the samples $\tilde{\vb*{s}}$ represent basis states in the rotated basis of the auxiliary space. However, this does not affect operators applied to the physical space, and therefore, the same estimator can be used.

\section{projected Imaginary Time Evolution (\pSR)}\label{sec:pite}
The energy gradients estimated in (time-dependent) variational Monte Carlo may be biased (see Ref.~\cite{sinibaldi2023unbiasing}). If we denote the elements of the computational basis by $x$, we have (we ignore the normalization for brevity of notation)
\begin{align}
    F_k &= \mel{\partial_k \Psi}{\hat{H}}{\Psi}, \\
    &= \sum_x \partial_k \Psi(x) \sum_{x'} \mel{x}{\hat{H}}{x'} \Psi(x'), \label{eq:_2} \\
    &= \mathbb{E}_{x \sim \abs{\Psi}^2} \left[ O_k^*(x) E_{\textrm{loc}}(x) \right] + B_k.
\end{align}
The first term is the standard estimator used in variational Monte Carlo while the contribution $B_k$ is typically neglected. This assumes that only configurations with finite Born probability $\abs{\Psi(x)}^2$ contribute to the gradients. The neglected bias term reads
\begin{align}
    B_k = \sum_{x | \Psi(x) = 0} \partial_k \Psi^*(x) \sum_{x'} \mel{x}{\hat{H}}{x'} \Psi(x'),
\end{align}
and contributes to the forces whenever the gradients do not vanish for configurations with zero Born probability. Although this bias term is often negligible in VMC applications, it is important for evolving certain states. An example of the latter is the infinite temperature state where many configurations $x$ with zero Born probabilities occur. Hence, this bias term explicitly indicates that we must consider the gradient of the wave function at basis states with zero probability, to obtain an unbiased estimate.

We aim to evolve a variational state for an imaginary time $\mathrm{d}\beta$ from $\ket{\psi_{\vb*{\theta}}(\beta)}$ into $\hat{\Pi}\ket{\psi_{\vb*{\theta}}(\beta)}$, with the imaginary-time propagator propagator $\hat{\Pi} = e^{-\frac{\mathrm{d}\beta}{2}\hat{H}_0\otimes\tilde{\mathds{1}}}$, and to represent (or compress) the resulting state with a new state $\ket{\psi_{\vb*{\eta}}(\beta)}$ on the variational manifold. To do so, our method consists of updating the variational parameters of the wave function so as to minimize the infidelity
\begin{equation}
    \vb*{\theta}(\beta+\mathrm{d} \beta) = \argmin_{\vb*{\eta}} \mathcal{I}\bigl(\ket{\psi_{\vb*{\eta}}}, \hat{\Pi}\ket{\psi_{\vb*{\theta}(\beta)}}\bigr), \label{eq:infidelity}
\end{equation}
In principle, the infidelity and gradients may be estimated by only sampling from $\lvert\psi_{\vb*{\eta}}\rvert^2$, as previously proposed for real-time evolution~\cite{donatella2022}. However, this approach is unable to lift the exact zeros of the wave function due to the bias problem introduced earlier. Furthermore, traditional stochastic reconfiguration cannot be employed since suffers from the same bias~\cite{sinibaldi2023unbiasing}. Therefore, we instead introduce the following new estimator for the infidelity $\mathcal{I} = 1-\mathcal{F}$ in Eq.~\eqref{eq:infidelity}:
\begin{align}
    \mathcal{F}\bigl(\ket{\psi}, \ket{\phi}\bigr)= \Re{A B}.
\end{align}
Here we use the simplifying notation $\ket{\psi} \equiv \ket{\psi_{\vb*{\eta}}(\beta)}$, $\ket{\phi} \equiv \hat{\Pi}\ket{\psi_{\vb*{\theta}}(\beta)}$, and
\begin{align}
    A &= \braket{\phi}{\psi}/\braket{\phi} = \underset{\vb*{S}\sim q}{\mathbb{E}} \left[w(\vb*{S};\phi)\frac{\psi(\vb*{S})}{\phi(\vb*{S})}\right]/\underset{\vb*{S}\sim q}{\mathbb{E}} \left[w(\vb*{S};\phi)\right],\nonumber\\
    B &= \braket{\psi}{\phi}/\braket{\psi} = \underset{\vb*{S}\sim q}{\mathbb{E}}\left[w(\vb*{S};\psi)\frac{\phi(\vb*{S})}{\psi(\vb*{S})}\right]/\underset{\vb*{S}\sim q}{\mathbb{E}} \left[w(\vb*{S};\psi)\right],
\end{align}
where $w(\vb*{S}; \varphi) := \lvert\varphi(\vb*{S})\rvert^2/q(\vb*{S})$ and where, as in the main text, $\vb*{S} := (\vb*{\sigma}, \tilde{\vb*{s}})$ was used to alleviate notations.

A crucial aspect of our method is that we introduced a non-parametrized distribution $q(\vb*{S})$, through which we perform importance sampling.
Furthermore, in our estimator, gradients of $\psi_{\vb*{\eta}}$ with respect to the parameters $\vb*{\eta}$ only need to be computed on the samples $\vb*{S}$ from this distribution $q$.
We introduce the following suitable prior distribution:
\begin{equation}
    q(\vb*{S}) =\mathcal{N} \sum_{\vb*{S}'} \lvert\braket{\vb*{S}'}{\iden}\rvert^2 f\bigl(D(\vb*{S}', \vb*{S})\bigr),
\end{equation}
where $\mathcal{N}$ is a normalization constant. Here, the convolutional kernel decays exponentially with the sample Hamming distances $D_H(\vb*{S}',\vb*{S}) := \sum_i 1-\braket{S'_i}{S_i}$ and
$f(D) = b^{-D}$ with hyperparameters $b$. The choice of Hamming distance is inspired by the fact that the state deviates from the identity state through powers of $\hat{H} \mathrm{d}t$ where $\hat{H}$ is local and where the importance of the $n$th power decays exponentially for small $\mathrm{d}t$.
Importance sampling proves crucial to optimize the infidelity to sufficiently low values. In terms of thermal state preparation, our approach allows us to reliably cool down the state from infinite temperature to remarkably low temperatures. 
Furthermore, we reduce the variance on the fidelity operator using the covariate technique in the recently introduced p-tVMC approach in Ref.~\cite{sinibaldi2023unbiasing}.

\section{Thermofield dynamics with tVMC}\label{sec:tvmc}
The real-time evolution of the thermofield state is governed by the Schrödinger equation. In variational space, this dynamical equation may be expressed implicitly as
\begin{equation}
    \dot{\vb*{\theta}} = \argmin_{\vb*{\eta}} \mathcal{D}(\ket{\psi_{\vb*{\theta} + \mathrm{d}t\vb*{\eta}}}, e^{-i\mathrm{d}t\hat{H}^\mathrm{th}}\ket{\psi_{\vb*{\theta}}}),
\end{equation}
where $\mathcal{D}$ is the Fubini-Study distance. To leading order in $\mathrm{d}t$, this leads to
\begin{equation}
    \vb{G}\dot{\vb*{\theta}} = -i\vb*{F}^\mathrm{th}, \label{eq:tdvp_thermal}
\end{equation}
where we introduced the quantum geometric tensor (QGT)
\begin{align}
    G_{k,k'} = \mathbb{E}\bigl[O_k^\star(\vb*{S}) O_{k'}(\vb*{S})\bigr] - \mathbb{E}\bigl[O_k^\star(\vb*{S})] \mathbb{E}\bigl[O_{k'}(\vb*{S})\bigr],
\end{align}
with log-derivatives $O_k$ and forces given by
\begin{gather}
    O_{k}(\vb*{S}) = \partial_{\theta_k}\ln\psi_{\vb*{\theta}}(\vb*{S}), \\
    F_k^\mathrm{th} = \mathbb{E}\bigl[ O_k^\star(\vb*{S}) E_{\mathrm{loc}}^\mathrm{th}(\vb*{S})\bigr]-\mathbb{E}\bigl[ O_k^\star(\vb*{S})\bigr] \mathbb{E}\bigl[E_{\mathrm{loc}}^\mathrm{th}(\vb*{S})\bigr].\label{eq:forces}
\end{gather}
In the above expectation values are implicitly taken with respect to the Born distribution $\lvert\psi_{\vb*{\theta}}\rvert^2$.

The local-energy estimator for the thermal Hamiltonian reads (see Eq.~\eqref{eq:physical_Oloc})
\begin{align}
    E_\mathrm{loc}^\mathrm{th}(\vb*{\sigma}, \tilde{\vb*{s}}) = &\sum_{\vb*{\sigma}'}\frac{\psi_{\vb*{\theta}}(\vb*{\sigma}', \tilde{\vb*{s}})}{\psi_{\vb*{\theta}}(\vb*{\sigma}, \tilde{\vb*{s}})}\mel{\vb*{\sigma}}{\hat{H}}{\tilde{\vb*{\sigma}}'} - \nonumber \\
    &\sum_{\tilde{\vb*{s}}'}\frac{\psi_{\vb*{\theta}}(\vb*{\sigma}, \tilde{\vb*{s}}')}{\psi_{\vb*{\theta}}(\vb*{\sigma}, \tilde{\vb*{s}})}\mel{\tilde{\vb*{s}}}{\tilde{H}}{\tilde{\vb*{s}}'}.
\end{align}

The above variational algorithm may be easily adapted in order to properly work with normalized \textit{Ansätze} by subtracting the variational energy from the thermal Hamiltonian $\hat{H}^{\mathrm{th}}\mapsto\hat{H}^{\mathrm{th}} - E^\mathrm{th}_{\vb*{\theta}}$. Furthermore, subtracting the average energy reduces the variance of the estimators in the case of unnormalized states.

\section{Thermal state architectures}\label{sec:architectures}
As mentioned in the main text, any thermal \textit{Ansatz} must be able to represent the infinite temperature state (the identity state in the thermofield formalism) exactly. Below, we introduce three methods to accomplish this goal, starting with a generic \textit{Ansatz}, eventually leading to the \TARNN and \TRBM ansatzes used in the results section of the main text.

\subsection{Generic neural-network architecture}
One can encode the structure of the identity state into a mean-field factor, thereby entirely lifting any restriction on the architecture of the model. Therefore, we realize that the identity state is a product state with maximally entangled physical-thermofield spin pairs. A possible construction therefore reads
\begin{equation}
    \ket{\Omega_{\vb*{\theta}}(\beta)} = \sum_{\vb*{\sigma},\tilde{\vb*{s}}} \psi_{\vb*{\theta}(\beta)}(\vb*{\sigma}, \tilde{\vb*{s}}) \bigotimes_{i=1}^N \sqrt{p_{\vb*{\theta}(\beta)}^\mathrm{MF}(\sigma_i, \tilde{s}_i)}\ket{\sigma_i, \tilde{s}_i},\label{eq:ntfs2}
\end{equation}
where $p_{\vb*{\theta}}^\mathrm{MF}$ denotes the mean-field factor. As discussed in the main text, there is some freedom in defining this state. However, in the $\hat{S}_z$ basis, the identity $\bigotimes_{i=1}^N\ket{\iden_i}$ can be represented exactly by the separable mean-field factor, by ensuring
\begin{equation}
    \psi_{\vb*{\theta}(\beta=0)}(\vb*{\sigma}, \tilde{\vb*{s}}) = \text{cst.},\quad
    p_{\vb*{\theta}(\beta=0)}^{\mathrm{MF}}(\sigma, \tilde{s}) \propto \prod\delta_{\sigma,\tilde{s}}
\end{equation}
at $\beta = 0$. In this scenario, the architecture of the neural thermofield state of Eq.~\eqref{eq:ntfs2} becomes clear. The mean-field separable factor captures the initial infinite-temperature structure while the neural network encodes both ``local'' correlations between physical and auxiliary spins and those between distant spins, which build up as the state is evolved in imaginary time to a finite inverse temperature $\beta$~\cite{bakshi2024high}. 

\subsection{\TARNN: ARNN architecture}

We introduce two possible \TARNN architectures: first, we start in the $S_z$ basis by turning the existing RNN architecture from Ref.~\cite{hibat-allah2020} into a thermal state. We point out that this ansatz would result in zero gradients in the infinite temperature state, and introduce a solution to this problem by representing the thermofield spins in a rotated basis.

\subsubsection{In the $(S_z, \tilde{S}_z)$ basis}

The NTFS in Eq.~\eqref{eq:ntfs1} can be implemented as a recurrent neural network in the $S_z$ basis with the following spin-pair autoregressive architecture:
\begin{align}
    \psi_{\vb*{\theta}}(\vb*{S}) &= \prod_{i=1}^N \psi^{(i)}_{\vb*{\theta}}(S_i|\vb*{S}_{<i}), \label{eq:rnn_factorized}
\end{align}
where we use the same notation $S_i = (\sigma_i, \tilde{s}_i)$ as in the main text. Furthermore, we decompose $\psi^{(i)}_{\vb*{\theta}} = \sqrt{p_{\vb*{\theta}}^\pidx{i}} e^{i\phi_{\vb*{\theta}}^\pidx{i}}$.
The Born distribution inherits this structure and samples can be drawn by iteratively producing pairs of physical-auxiliary spin configurations from the first to the last site without the need for a Markov chain. The mean-field factor at every site is then simply chosen as $\sqrt{p^{\mathrm{MF}}}(\sigma, \tilde{s}) = \alpha^2\delta_{\sigma, s}$, where $\alpha$ is a real variational parameter. By design, this factor is compatible with the autoregressive property. In sum, the thermofield identity state can be represented purely through a careful design of the output layers at every site.

\subsubsection{In the $(S_z, \tilde{S}_x)$ basis\label{SM:A2}}

The above implementation suffers from zero gradients at the initial $\beta = 0$ and must thus be first evolved through the \pSR method described in the main text. This stems from the presence of exact zeros in the identity-state wave function. Alternatively, one can perform a Hadamard rotation $\hat{\mathds{1}}\otimes(\tilde{\sigma}^x + \tilde{\sigma}^z)/\sqrt{2}$ of the local computational basis. This transformation only acts on the thermofield space and therefore has no incidence on imaginary-time evolution, which only involves the action of physical operators.

Let us consider
\begin{equation}
    \ket{0} = \ket{0\tilde{+}},\quad
    \ket{1} = \ket{0\tilde{-}},\quad
    \ket{2} = \ket{1\tilde{+}},\quad
    \ket{3} = \ket{1\tilde{-}} \label{eq:basis_4};
\end{equation}
as the computational basis of this local space of physical-thermofield pairs. Then, one needs to impose
\begin{align}
    \braket{0, \tilde{1}}{\psi_{\vb*{\theta}(0)}} \propto \braket{0,\tilde{+}}{\psi_{\vb*{\theta}(0)}} - \braket{0\tilde{-}}{\psi_{\vb*{\theta}(0)}} &= 0 , \\
    \braket{1, \tilde{0}}{\psi_{\vb*{\theta}(0)}} \propto \braket{1,\tilde{+}}{\psi_{\vb*{\theta}(0)}} + \braket{1\tilde{-}}{\psi_{\vb*{\theta}(0)}} &= 0 ;
\end{align}
and
\begin{align}
    \braket{0, \tilde{0}}{\psi_{\vb*{\theta}(0)}} \propto \braket{0,\tilde{+}}{\psi_{\vb*{\theta}(0)}} + \braket{0\tilde{-}}{\psi_{\vb*{\theta}(0)}} &= C , \\
    \braket{1, \tilde{1}}{\psi_{\vb*{\theta}(0)}} \propto \braket{1,\tilde{+}}{\psi_{\vb*{\theta}(0)}} - \braket{1\tilde{-}}{\psi_{\vb*{\theta}(0)}} &= C .
\end{align}
In the above-defined basis, we investigate how such a state can be constructed in the form of Eq.~\eqref{eq:rnn_factorized}. First, we defined the normalized probability $p^{(i)}$ based on the unnormalized probability vector $\tilde{\vb*{p}}^{(i)}$ in the local basis,
\begin{equation}
\begin{gathered}
    \ln p_{\vb*{\theta}}^\pidx{i} = \mathrm{L}\Sigma\mathrm{E}(\ln \tilde{\vb*{p}}_{\vb*{\theta}}^\pidx{i}),
\end{gathered}
\end{equation}
where the logsumexp ($\mathrm{L}\Sigma\mathrm{E}$) sums over the 4-dimensional local Hilbert space basis in Eq.~\eqref{eq:basis_4}.
We now aim to construct the unnormalized probabilities and phases such that
\begin{align}
    \ln \tilde{\vb*{p}}_{\vb*{\theta}}^\pidx{i} &= [\gamma_i, \gamma_i, \gamma_i, \gamma_i],\\
    \tilde{\vb*{\phi}}_{\vb*{\theta}}^\pidx{i} &= [\delta_i, \delta_i, \delta_i, \delta_i\pm\pi].
\end{align}
In an autoregressive NQS, the amplitudes and phases are obtained as
\begin{equation}
\ln \tilde{\vb*{p}}_{\vb*{\theta}}^\pidx{i} = \vb{W}_{\mathrm{amp}}\vb*{h}^\pidx{i} + \vb*{b}_\mathrm{amp},    
\end{equation}
and
\begin{equation}
    \vb*{\phi}_{\vb*{\theta}}^\pidx{i} = \pi  \left(\vb{W}_{\mathrm{ph}}\vb*{h}^\pidx{i} + \vb*{b}_\mathrm{ph}\right),
\end{equation}
where
\begin{equation}
    \vb*{h}^{\pidx{i}} = \mathrm{RNN}\bigl(\lbrace S_j\rbrace_{j\to i}; \lbrace\vb*{h}^{\pidx{j}}\rbrace_{j\rightarrow i}\bigr)
\end{equation}
is the hidden vector at site $i$. Here, $\vb*{h}^\pidx{i}\in \mathbb{R}^{P_h}$, $\vb*{b}_{\mathrm{amp}}, \vb*{b}_{\mathrm{ph}} \in \mathbb{R}^{d^2}$, $\vb{W}_{\mathrm{amp}}, \vb{W}_{\mathrm{ph}} \in \mathbb{R}^{d^2\times P_h}$. For reference, the RNN cell used in the main text is an LSTM cell.

In practice, the above identity-state requirements can be fulfilled by generating initial random parameters as
\begin{equation}
\begin{gathered}
    \vb{W}_{\mathrm{amp}}(\beta=0) = \vb{W}_{\mathrm{ph}}(\beta=0) = \vb{0}, \\
    \vb*{b}_\mathrm{amp}(\beta=0) = x\vb*{1},\quad x \sim \mathcal{N}(0, \sigma), \\
    % \vb{b}_\mathrm{ph}(\beta=0) = (y, y, y + \pi, y),\quad y \sim \mathcal{N}(0, \sigma);
    \vb{b}_\mathrm{ph}(\beta=0) = (y, y, y, y-1),\quad y \sim \mathcal{N}(0, \sigma);
\end{gathered}
\end{equation}
where we set $\sigma=0.01$.

This neural quantum state does not suffer from zero initial gradients and can thus be evolved in imaginary time with stochastic reconfiguration from $\beta = 0$. However, when performing real-time evolution, the thermal Hamiltonian must be partially rotated $\hat{H}^\mathrm{th} \mapsto (\hat{\mathds{1}}\otimes \tilde{\mathrm{H}}\mathrm{ad}_N) \hat{H}^\mathrm{th} (\hat{\mathds{1}}\otimes \tilde{\mathrm{H}}\mathrm{ad}_N)$, where
\begin{equation}
    \tilde{\mathrm{H}}\mathrm{ad}_N = \bigotimes_{i=1}^N(\tilde{\sigma}_i^x + \tilde{\sigma}_i^z)/\sqrt{2}
\end{equation}
denotes the Hadamard rotation operated on the auxiliary degrees of freedom.

% For the RBMO architecture, we refer to the Supplemental Material~\cite{Note1}.

\subsection{RBMO architecture}
An RBM thermal state can be obtained as
\begin{align}
    \ln\psi_{\vb*{\theta}}(\vb*{\sigma}, \tilde{\vb*{s}}) &= \sum_i (a_i\sigma_i+a'_i \tilde{s}_i)\label{eq:rbm}\\
    &+\sum_m\ln\cosh\Bigl(b_m + \sum_j W_{mj}\sigma_j + \sum_j W'_{mj}\tilde{s}_j\Bigr).\nonumber
\end{align}

In order to produce the initial identity state, the parameters are initialized as
\begin{equation}
    a_i = a'_i = b_i = 0,\quad
    W_{ij} = -W'_{ij} = \frac{i\pi}{4} \delta_{i,j}.
\end{equation}
This yields $\ket{\psi_{\vb*{\theta}}(\beta=0)} \propto \bigotimes_i(\ket{\smash{\uparrow\tilde{\uparrow}}}_i + \ket{\smash{\downarrow\tilde{\downarrow}}}_i)$, \textit{i.e.} the identity state. This initialization is slightly different from that used in Ref.~\cite{nomura2021} and allows working in the $S_z$ basis for both the physical and auxiliary space, which considerably simplifies real-time dynamics. Contrary to the previous subsection, this architecture suffers from exact zero gradients at $\beta=0$ and hence requires the initial cooling down with p-ITE.

\section{Thermal state limits}\label{sec:thermal_limits}
\begin{figure*}[htb]
    \centering
    \includegraphics[width=0.67\textwidth]{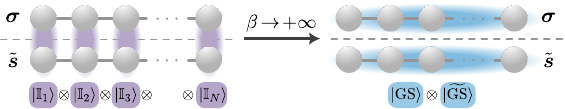}
    \caption{Schematic illustration of the thermofield representation of the quantum state in the extreme temperature limits. In this picture, the maximally mixed state ($T=+\infty$) corresponds to a product of locally pairwise maximally entangled Bell states; the pure state ($T=0$) corresponds to two copies of the ground-state wave function, whose correlations may be faithfully encoded by a neural-network \emph{Ansatz}.}
    \label{fig:picture_limits}
\end{figure*}

We have demonstrated how the initial infinite-temperature state can be prepared in various frameworks. In the asymptotic limit $\beta \to +\infty$, the thermofield state factorizes into a tensor product of the ground state on the physical space $\ket{\mathrm{GS}}$ and the thermofield space $\ket{\smash{\widetilde{\mathrm{GS}}}}$, as illustrated in Fig.~\ref{fig:picture_limits}. Provided an RBM with a hidden-spin density of $\alpha$ can capture the ground-state of the zero-temperature system, it is straightforward to show that the above factorized quantum state $\ket{\Psi} \otimes \ket{\smash{\tilde{\Psi}}}$ can also be captured by the RBMO with a hidden-spin density of $2\alpha$. Indeed, upon setting the following constraints:
\begin{align}
    W_{m, j} &= 0, \forall  m > N \\
    W_{m, j}' &= 0, \forall m \leq N
\end{align}

one has
\begin{align}
    \ln\psi_{\vb*{\theta}}(\vb*{\sigma}, \tilde{\vb*{s}}) &= \left[\sum_i a_i\sigma_i +\sum_{m=1}^N\ln\cosh\Bigl(b_m + \sum_j W_{mj}\sigma_j \Bigr)\right] \\
    &+ \left[\sum_i a'_i \tilde{s}_i +\sum_{m=N+1}^{2N}\ln\cosh\Bigl(b_m + \sum_j W'_{mj}\tilde{s}_j\Bigr)\right].\nonumber
\end{align}
which explicitly factorizes into a product of the same state if 
\begin{align}
    W_{m, j}' &= W_{m-N, j},\; \forall m > N; \\
    a_i' &= a_i,\; \forall i;\\
    b_m &= b_{m-N},\; \forall m > N.
\end{align}

\section{Thermofield algebra}\label{sec:thermofield_algebra}
The thermofield spin-operator $\tilde{O}$ commutes with operators on the physical Hilbert space, and has the matrix elements
\begin{align}
    \mel{\tilde{s}}{\tilde{O}}{\tilde{s}'} &= \braket{\smash{\tilde{s}}}{\smash{\widetilde{O s'}}} \\
    &= \bra{\tilde{s}} \left( \sum_{\lambda=1}^{\dim \mathcal{H}} \mel{\lambda}{\hat{O}}{\tilde{s}'}^* \ket{\smash{\tilde{\lambda}}} \right) \\
    &= \mel{s}{\hat{O}}{s'}^*,
\end{align}
where we used Eq.~(S2) from the Supplemental Material~\cite{Note1}. In other words, we have $\tilde{O} \to \hat{O}^*$, or $\tilde{X} \to X$, $\tilde{Y} \to -Y$ and $\tilde{Z} \to Z$. See also the Thermo-SU(2) algebra discussion in Ref.~\cite{petronilo2021simulating}.

\section{Benchmark on 1D spin chain}\label{sec:chain_benchmark}
\begin{figure*}[tbh]
    \centering
    \includegraphics[width=0.85\textwidth]{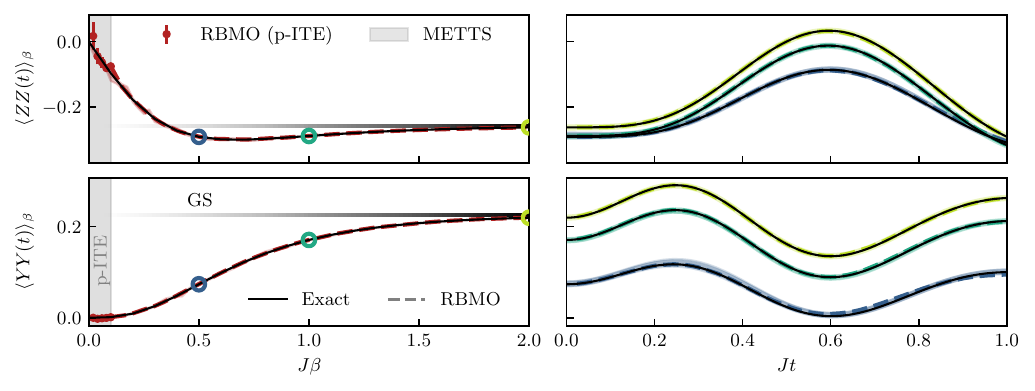}
    \caption{Thermal observables evolution as a function of $\beta$ (left) and real time $t$ (right) for a spin chain of length $10$. The thin black line indicates the exact values. METTS predictions (with $3\sigma$ bands, 10k samples) and ground-state (GS) values ($\beta \to \infty$) are included for comparison. Observables in the \pSR temperature regime are estimated via standard local Monte Carlo sampling and thus have higher variance than the sampling scheme used in the actual \pSR propagation. Operators are evaluated on neighboring sites.
    }
    \label{fig:N10_D1}
\end{figure*}

In the main text, we show predictions of time-dependent thermal observables on a $4{\times}4$ lattice and compare with METTS predictions to validate our approach. Here we also show predictions for a periodic spin chain of $10$ sites and compare with predictions from exact diagonalization (using the exact matrix exponentials in the time evolution and using the full density matrix). The results are shown in Fig.~\ref{fig:N10_D1} for the observables $ZZ$ and $YY$ as in the main text. The thermal state is prepared for the Hamiltonian $\hat{H}_0 = \hat{H}(2h_c, 0)$ with $h_c=1$ in 1D.
These results again suggest an excellent agreement over a wide temperature range, with the most accurate predictions at lower temperatures.

\section{Comparison of $4{\times}4$ vs $6{\times}6$}\label{sec:comparison_4x4_and_6x6}
\begin{figure*}[htb]
    \centering
    \includegraphics[width=.5\textwidth]{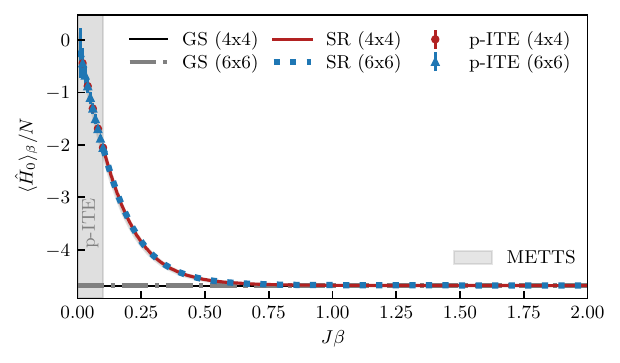}
    \caption{Energy per site as a function of inverse temperature $\beta$ for the $4{\times}4$ and $6{\times}6$ spin torus. Labels are as in Fig.~\ref{fig:N4_D2}.}
    \label{fig:energy_overlap_2d}
\end{figure*}

Since we do not have exact predictions available for the time evolution of thermal observables of a $6{\times}6$ lattice, we validate our predictions by matching the energy per site with the $4{\times}4$ lattice in Fig.~\ref{fig:energy_overlap_2d}. Although the results of both lattice configurations were produced with different models (see the numerical details section), we obtain an excellent overlap of both thermal energy densities, indicating that our $6{\times}6$ density matrices are trustworthy over the full temperature range.

\newpage
\FloatBarrier
\newpage

%\onecolumngrid

%\widetext
\clearpage
\begin{center}
\textbf{\large Supplemental Material to ``\textit{Real-time quantum dynamics of thermal states with neural thermofields}"}
\end{center}
%%%%%%%%%% Merge with supplemental materials %%%%%%%%%%
%%%%%%%%%% Prefix a "S" to all equations, figures, tables and reset the counter %%%%%%%%%%
%\setcounter{equation}{0}
%\setcounter{figure}{0}
%\setcounter{table}{0}
%\setcounter{page}{1}
\makeatletter
%\renewcommand{\theequation}{S\arabic{equation}}
%\renewcommand{\thefigure}{S\arabic{figure}}
%\renewcommand{\bibnumfmt}[1]{[S#1]}
%\renewcommand{\citenumfont}[1]{S#1}
%%%%%%%%%% Prefix a "S" to all equations, figures, tables and reset the counter %%%%%%%%%%

%\section*{Supplemental Material to ``\textit{Real-time quantum dynamics of thermal states with neural thermofields}"}

 \renewcommand{\thefigure}{S\arabic{figure}}
 \renewcommand{\theequation}{S\arabic{equation}}
 \renewcommand{\thesection}{S\arabic{section}}
 \setcounter{equation}{0}
 \setcounter{figure}{0}
 \setcounter{section}{0}
   \setcounter{subsection}{0}
%\titleformat{\section}{\large\bfseries}{\thesection .}%{0.5em}{}
%\titleformat{\section}{\bfseries}{\thesection.}{0.5em}{}

%\tableofcontents

\section{Connection to the density-matrix approach}\label{sec:connection_to_dm}

The connection is direct through Liouville space, see Ref. \cite{arimitsu1985,gyamfi2020}. Let $\mathcal{O}$ denote the space of operators acting on the physical Hilbert space $\mathcal{H}$, the action of any superoperator onto any element of $\mathcal{O}$ can be mapped into the action of an operator onto a Liouville space of \emph{superkets} $\mathcal{L}$. The isomorphism between $\mathcal{O}$ and $\mathcal{L}$ is given by the action of the bra-flipping linear superoperator $\mathcal{T}$:
\begin{equation}
    \mathcal{T}\bigl[\ketbra{\psi}{\phi}\bigr] = \ket{\psi}\otimes\widetilde{\ket{\phi}},
\end{equation}
where one has
\begin{equation}
    \widetilde{\ket{\phi}} = \sum_{n=1}^{\dim \mathcal{H}} \braket{n}{\phi}^\star\ket{\tilde{n}}. \label{eq:tilde_decomposition}
\end{equation}
This specific behavior of the tilding operation ensures that $\mathcal{T}$ be an isomorphism:
\begin{equation}
    \bigl\langle \mathcal{T}\hat{A}, \mathcal{T}\hat{B}\bigr\rangle_{\mathcal{L}} \overset{!}{=} \Tr\bigl[\hat{A}^\dagger\hat{B}\bigr] \equiv \bigl\langle \hat{A}, \hat{B}\bigr\rangle_{\mathcal{O}}.
\end{equation}

The transformation of superoperators into operators acting on Liouville space can be derived from the above:
\begin{equation}
    \mathcal{T}\bigl[\hat{A}\ketbra{\psi}{\phi}\hat{B}\bigr] = \hat{A}\ket{\psi}\otimes\widetilde{\ket{\smash{\hat{B}\phi}}} = \hat{A}\otimes\tilde{B}\ket{\smash{\psi, \tilde{\phi}}},
\end{equation}
which induces some properties on the action of tilde on operators, for instance
\begin{align}
    (\widetilde{\alpha\hat{A} + \beta\hat{B}})\ket{\smash{\tilde{\phi}}} &= \widetilde{\ket{\smash{(\alpha\hat{A} + \beta\hat{B})\phi}}}\nonumber\\
    &= \sum_{n=1}^{\dim \mathcal{H}} \Bigl(\alpha^\star\mel{n}{\hat{A}}{\phi}^\star\ket{\tilde{n}} + \beta^\star\mel{n}{\hat{B}}{\phi}^\star\ket{\tilde{n}}\Bigr)\nonumber\\
    &= (\alpha^\star\tilde{A} + \beta^\star\tilde{B}) \ket{\smash{\tilde{\phi}}}.
\end{align}
From these properties, one can translate the action of superoperators from operator space into Liouville space. For instance,
\begin{align}
    \mathcal{T}\Bigl[ [\hat{H}, \placeholderdot] \ketbra{\psi}{\phi} \Bigr] &= \mathcal{T}\Bigl[ \ketbra{\smash{\hat{H}\psi}}{\phi} \Bigr] - \mathcal{T}\Bigl[ \ketbra{\psi}{\smash{\hat{H}\phi}} \Bigr]\nonumber\\
    &= \bigl(\hat{H}\otimes\tilde{\mathds{1}} - \hat{\mathds{1}}\otimes\tilde{H}\bigr)\ket{\smash{\psi, \tilde{\phi}}},
\end{align}
where it becomes clear that tilde space simply accounts for the right action of superoperators on any operator belonging to $\mathcal{O}$. This connection has also been laid out in detail recently in Ref.~\cite{kothe2023liouville}.

\section{Scalability}\label{sec:scalability}
We will discuss the scalability of our approach in terms of the number of physical spins $N$ for simplicity. The asymptotic scaling behavior is the same when the doubled space is taken into account (up to prefactors).
Time-evolving a neural quantum state requires the computation of local energies and the log-derivatives of the model on a set of $N_s$ Monte Carlo samples. Computing the local energy is $\order{N}$ per sample for lattice spin Hamiltonians. Obtaining MC samples requires $\order{N N_{\textrm{sweeps}}}$ evaluations of the model, where $N_{\textrm{sweeps}}$ is a user-defined thinning factor to decorrelate subsequent samples ($N_{\textrm{sweeps}}=1$ for ARNNO), and depends on the correlation time. 
The number of parameters $N_P$ in RBMO in general scales as $\order{\alpha N^2}$ with $\alpha N$ the number of hidden nodes. We take $\alpha=1$ everywhere. The number of parameters in the RBMO can be reduced to $\order{\alpha N}$ when using the translation-invariant form in Ref.~\cite{carleo2017}. For the ARNNO model, the number of parameters is independent of the system size and the computation of the probability amplitudes scales linearly in the system size. The computational cost of a forward pass to compute the probability amplitudes is $\order{N^2}$ and $\order{N}$ for the ARNNO model. The backward pass has the same complexity as the forward pass in both models due to the use of automatic differentiation and backpropagation.
The most computationally expensive part for the RBMO is solving the TDVP equations. Hereby, we invert the QGT using SVD, for which the computational burden scales as $\order{N_P^3}$. 

\section{Numerical details}\label{sec:numerical_details}
We use time steps $\mathrm{d}\beta \in \{0.005, 0.01\}/2$ for the initial \pSR evolution up to $\beta = 0.1$ ($\mathcal{C}_1$), and $\mathrm{d}\beta \in \{10^{-3}, 10^{-4}\}/2$ for the second imaginary time evolution using SR ($\mathcal{C}_2$). We use the Runge-Kutta 2 and 4 for the SR evolution with RBM and \TARNN respectively. For numerical stability, we use the SVD regularization defined in Ref.~\cite{medvidovic2022towards} with an absolute cutoff $\texttt{atol} \in \{10^{-7}, 10^{-8}\}$ for the imaginary and real-time evolution. For the \pSR, SR and tVMC, we use around $16$k, $64$k and $500$k samples respectively. For the $6{\times}6$ lattice, we use an RBM that is symmetrized over the translation group~\cite{carleo2017}. We take the hidden dimension fraction of the RBM to be $\alpha=1$ (i.e.\ $N_h = \alpha N$) everywhere, and $N_h = 8$ and $1$ layer for the ARNNO. For clarity, the data in Figs. 1, 2, 3 and 4 in the main text were smoothed with a moving average window of size 50.

\section{Calculating thermodynamic observables with METTS}

As a validation of our results on the spin chain of length $10$ and 2D lattice of $4{\times}4$ spins, we compute thermodynamic observables using the minimally entangled thermal states algorithm (METTS)~\cite{white2009,bruognolo2017,stoudenmire2010}.
For an observable $O$ and inverse temperature $\beta$, the METTS algorithm aims to generate a set of samples $\{O_m\}_m$ whose average approximate the 
thermal ensemble average
\begin{equation*}
    \expval{\smash{\hat{O}}}_\beta = \frac{\Tr(e^{-\beta \hat{H}} \hat{O})}{\Tr(e^{-\beta H})} = \lim_{M\to+\infty} \frac{1}{M} \sum_{m=1}^{M} O_m.
\end{equation*}

Here, the samples $O_m$ are generated via a Markov-chain process as follows:
\begin{enumerate}
    \item Start from a classical basis state $\ket{n}$.
    \item Evolve it up to imaginary time $t=-i\beta/2$
    \begin{equation*}
        \ket{\phi_n(\beta/2)} \propto e^{-\beta \hat{H}/2 }\ket{n}.
    \end{equation*}
    For this, we use exact state vectors and the Tsitouras 5/4 Runge-Kutta integration method (free 4th order interpolant)~\cite{rackauckas2017differentialequations} to solve the differential equation ($\tau = it$)
    \begin{equation*}
        \fder{\ket{\phi_n (\tau)}}{\tau} = - \left(\hat{H}_0 - E_\tau\right) \ket{\phi_n (\tau)} \,,
    \end{equation*}
    where $E_\tau = \mel{\phi_n (\tau)}{\hat{H}_0}{\phi_n (\tau)}/\braket{\phi_n (\tau)}$.
    \item Evaluate the observable to obtain the sample
    \begin{equation*}
        O_m \leftarrow \frac{\mel{\phi_n(\beta/2)}{\hat{O}}{\phi_n(\beta/2)}}{\braket{\phi_n(\beta/2)}}.
    \end{equation*}
    \item Sample from the Born distribution of $\ket{\phi_n(\beta/2)}$ in a chosen basis to obtain the next seed basis state and start again from Step 1.
\end{enumerate}
To reduce the correlations in the Markov chains between subsequent loops, we alternate between measurements in the  $Z$ and $X$ basis in Step 4 using Hadamard rotations. This proves crucial, especially at high temperatures. We discard the first $100$ samples of each Markov chain in order to ensure they are properly thermalized.

\end{document}